\documentstyle[aps,eqsecnum,epsf,epsfig]{revtex}

\begin{document}

\baselineskip=16pt

\title{Scenario for Ultrarelativistic Nuclear Collisions: \\
  IV.~ Effective quark mass at the early stage. } 
\author{ A.  Makhlin and  E. Surdutovich}
\address{Department of Physics and Astronomy, Wayne State University, 
Detroit, MI 48202}
\date{July 26, 2000}
\maketitle

\begin{abstract}

Using the framework of wedge dynamics, we compute the effective transverse
mass of a soft quark mode propagating in the expanding background of hard
quarks and gluons created at the earliest time of the collision.  We
discover that the wedge dynamics  does not require any external infrared
or collinear cut-off. The effective mass is produced mainly due to the
forward quark-quark scattering mediated by the longitudinal (in sense of
Gauss' law) magnetic fields. Contribution of the radiation field is
parametrically suppressed.

\end{abstract}

\section{Introduction }
\label{sec:S1}

In the first paper of this cycle \cite{tev} ( further quoted as paper
[I]), we formulated a program that might result in a theory of
ultrarelativistic nuclear collisions which is free from collinear problems
and naturally establishes the infrared boundary for the space of ``final''
states  at the very early stage of the collision ($\leq 1fm$).  We have
demonstrated that even at very early times (much less than is required for
any kinetic process to develop), the collective interactions in a dense
system provide the final states of the QCD evolution with finite
dynamically generated masses that  shield mass singularities in the
evolution equations.\footnote{ The idea that screening effects should be
taken into account at the early kinetic stage of a collision  has been
articulated earlier and with different motivations by Shuryak and Xiong
\cite{Xiong} and by Eskola, Muller and X.-N. Wang \cite{Eskola}}. It was
shown also that the null-plane dynamics are incapable of describing  local
screening effects, because any type of kinetics is frozen on the light
cone. It was suggested, that a more adequate approach requires the change
of the global Hamiltonian dynamics which is used for the field-theory 
description of  nuclear collisions.  We proposed the so-called {\em wedge
dynamics} which employs the proper time $\tau$ measured from the first
touch of the Lorenz-contracted nuclei as the Hamiltonian time. Our initial
estimates in paper [I] were very qualitative. In two consecutive papers
\cite{gqm,wdg}  (further quoted as papers [II] and [III]), we have
studied, in detail, the space of states of wedge dynamics. In paper [II],
we extended the qualitative analysis of scalar fields initiated in [I],
and found that for the charged fields, the early-time evolution of the
wave function is accompanied by a gradual rearrangement of the charge
distribution, starting from its almost uniform spread along the light cone
at $\tau\to 0$, and up to a narrow wave packet with a well defined
rapidity at later times. We have shown that this re-distribution of the
charge leads to currents in the rapidity direction and that these currents
are the largest at the earliest $\tau$. The magnetic fields generated by
these currents can be  responsible for the interactions between the
currents at the earliest moments of the QCD evolution. In paper [II], we
studied the states of fermions in wedge dynamics and found the fermion
field correlators that are used below for the calculation of the quark
self-energy in the expanding system. In paper [III], we addressed the
issue of gauge fields in wedge dynamics. Several important problems were
solved there. The natural gauge condition of wedge dynamics, $A^\tau =0$,
was proved to be completely fixed (at the level of perturbation theory).
The second (technically nearly most difficult) problem solved in paper
[III], was the separation of the longitudinal (i.e., governed by Gauss'
law) field and the field of radiation. In that paper, we also quantized
the gauge field in the scope of wedge dynamics and explicitly found the
Wightman functions and retarded propagator of the gluon field which are
used in this paper for the practical calculation of the fermion
self-energy.

Our decision to begin the exploration of potentialities of the wedge
dynamics with the  computation of quark self-energy is motivated only by
technical reasons. The gluon propagator of wedge dynamics is a very
complicated function, and we preferred to start with the computation of
the fermion loop which has only one gluon correlator in it. We hope that
the possibility of a technical simplification (compared to what we had to
start with) discovered in this paper, will allow us to address a more
important problem of the gluon self-energy in a reasonably economic way.

In the course of this study, we employ a single  heuristic assumption
(supported by the analysis of paper [II]) that the field states with large
transverse momentum, even at very early times, may be associated with the
localized particles and thus can be described by the distribution with
respect to the rapidity and transverse momentum.  Our strategy of looking
for the leading contributions, as well as all our approximations, in the
calculation of the material part of the quark self-energy are based on
this assumption. If it appears incorrect, then  it is most likely that the
quark-gluon matter created in the collision of  two nuclei never and in no
approximation can be considered as a system of nearly free and weakly
interacting field states.

\section{Fermion retarded self-energy }
\label{sec:S3}

In order to find the normal modes of the  quark field in the expanding
quark-gluon system , we are going to solve the Dirac equation with the
radiative corrections, which can be derived as a projection of the
Schwinger-Dyson equation for the retarded quark propagator onto the
one-particle initial state. For the quark field without Lagrangian mass,
this equation reads
\begin{eqnarray}  
i \gamma^\mu (x_1) \nabla_\mu (x_1) \psi(x_1)= 
\int d^4 x_2 \Sigma_{[ret]}(x_1,x_2) \psi(x_2)~.
\label{eq:E3.0a}\end{eqnarray}    
The covariant derivative $\nabla_\mu (x)$ of the spinor field in the
curvilinear coordinates of the wedge dynamics includes the spin connection
and it was  found explicitly in paper [II]. For all calculations below, we
employ the mixed representation which is the most profitable in heavy-ion
problems. We are looking for the radiative corrections to the wave
function with a given transverse momentum $\vec{p_t}$ and rapidity
$\theta$ with the expectation that within the rapidity plateau nothing
will depend on $\theta$. However, we cannot totally eliminate the
coordinate $\eta$ from the theory. We have to keep it explicitly, since 
the problem of the expanding field system cannot be reduced to (2+1)
dimensions. In its expanded form, Eq.~(\ref{eq:E3.0a})  reads,
\begin{eqnarray}
\bigg[i\gamma^0\big({\partial\over\partial\tau_1}+{1\over 2\tau_1}\big) 
+{i\gamma^3\over\tau_1}{\partial\over\partial\eta_1} -p_r\gamma^r\bigg]
\psi (p_t;\tau_1,\eta_1)=
\int_{0}^{\tau_1}d\tau_2\int_{-\infty}^{\infty}\tau_2 d\eta_2
\Sigma_{[ret]}(p_t;\tau_1,\tau_2;\eta_1-\eta_2)\psi (p_t;\tau_2,\eta_2)~.
\label{eq:E3.0b}
\end{eqnarray}
The retarded self-energy is an object that naturally emerges in the
Schwinger-Dyson equation for the retarded propagator in Keldysh-Schwinger
formalism \cite{Keld}.  Below,  we employ its modified form developed
earlier  with the view of application to the inclusive and transient
processes. We employ the notation used in 
Refs.~\cite{QFK,QGD,tev}.\footnote{The indices  of the field correlators
with the Keldysh contour ordering of the field operators (like $G_{[AB]}$)
as well as the labels of their linear combinations (like $G_{[ret]}$) are
placed in square brackets.} In this notation, the one-loop retarded
fermion self-energy  in coordinate form is
\begin{eqnarray}
  \Sigma_{[ret]}(x_1,x_2)={ig^{2}\over 2}
  [t^{a} \gamma^{\mu} G_{[ret]}(x_1,x_2)t^{b}\gamma^{\lambda}
  D^{ba}_{[1]\lambda\mu}(x_2,x_1)+t^{a} \gamma^{\mu} G_{[1]}(x_1,x_2)
   t^{b}\gamma^{\lambda} D^{ba}_{[adv]\lambda\mu}(x_2,x_1)]~.  
\label{eq:E3.1}
\end{eqnarray}
The two subprocesses that contribute this self-energy are depicted below.
\begin{figure}[htb]
\begin{center}
\mbox{ 
\psfig{file=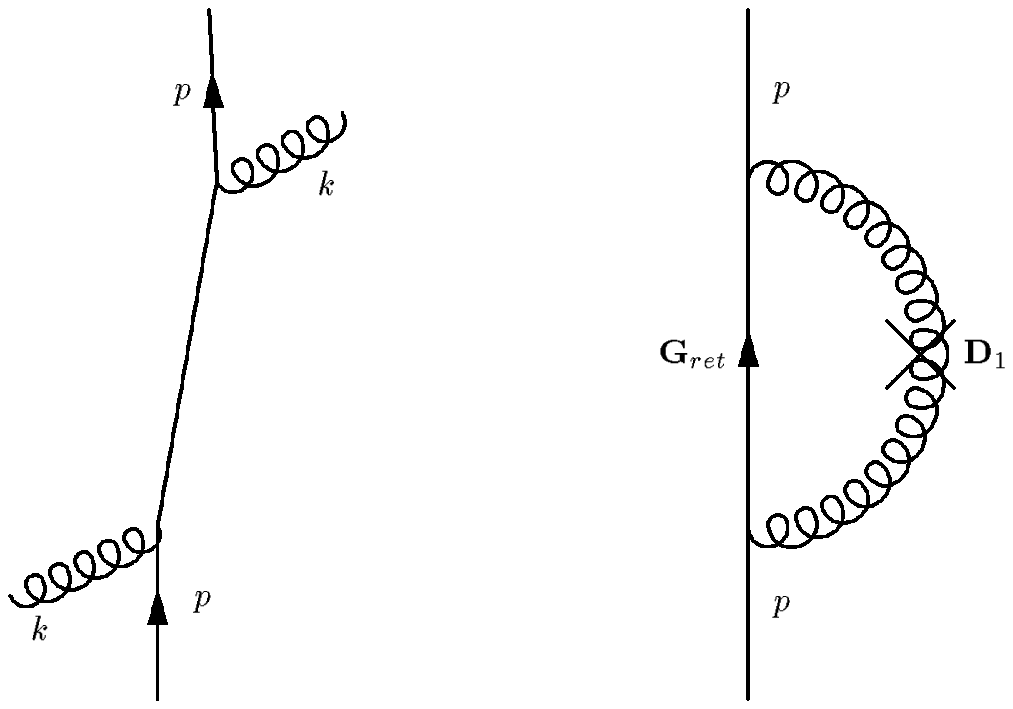,height=2.in,bb=130 460 440 675}
\hspace{2.cm}
\psfig{file=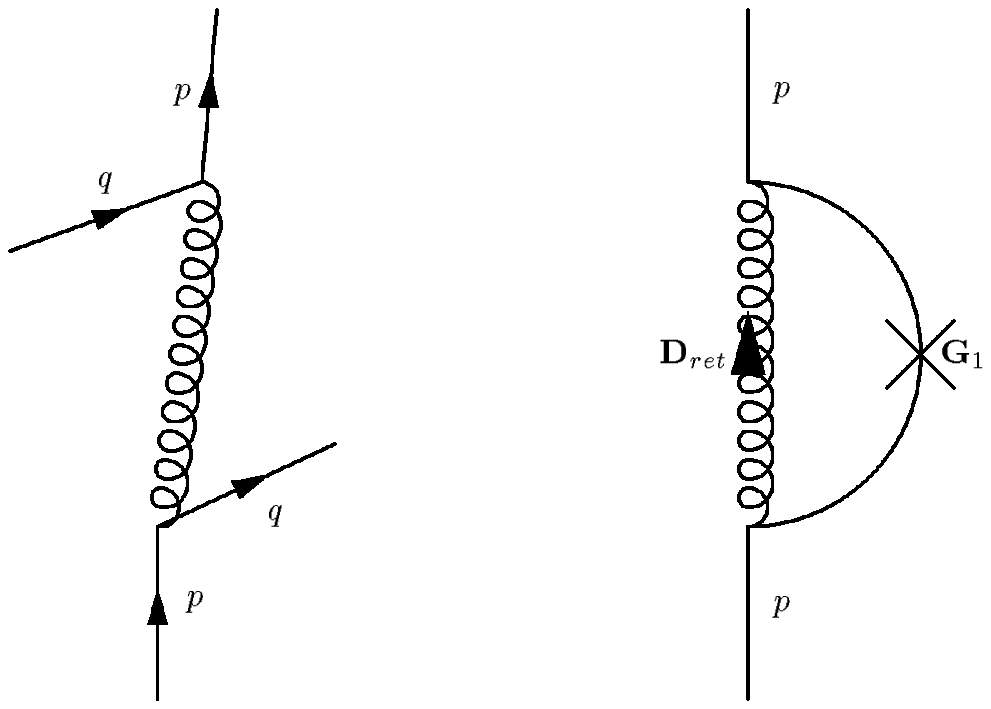,height=2.in,bb=140 460 430 670}
}
\end{center}
\caption{The retarded forward scattering amplitude is contributed by two
subprocesses, $qg\to qg$ and $qq\to qq$.}
\label{fig:fig4}
\end{figure}

The retarded and advanced quark and gluon propagators $G_{[ret]}$ and
$D^{lm}_{[adv]}$ were found in papers [II] and [III] of this cycle
and are connected with the commutators  $G_{[0]}$ and $D_{[0]}$,
\begin{eqnarray}
G_{[ret]}(x_1,x_2)=\theta(\tau_1-\tau_2)G_{[0]}(x_1,x_2),~~
D^{lm}_{[adv]}(x_2,x_1)=-\theta(\tau_1-\tau_2)D^{lm}_{[0]}(x_2,x_1)
+D^{lm}_{L}(x_2,x_1)~,
\label{eq:E3.2}
\end{eqnarray}
where $D^{lm}_{L}(x_2,x_1)$ is the longitudinal part of the gluon
propagator (governed by Gauss' law), and it enters in Eq.~(\ref{eq:E3.2})
in such a way that the condition $D_{[ret]}-D_{[adv]}=D_{[0]}$ is
satisfied and the non-causal longitudinal part of the propagator does not
violate the causal properties of the commutator  $D_{[0]}$. The
correlators $G_{[1]}$ and $D_{[1]}$  include densities of vacuum states as
well as the  information about the occupation numbers (phase-space
population). Eventually, we shall prove that an approximation of the
boost-invariance (infinite rapidity plateau) is not corrupted by any kind
of cut-offs (the vacuum part never is). Therefore, all correlators ($G$,
$D$, and $\Sigma$) will depend on two times $\tau_1$ and $\tau_2$
separately, the difference of rapidities $\eta=\eta_1-\eta_2$, and the
difference  ${\vec r} ={\vec r_1}-{\vec r_2}$ of distances in $xy$ plane.
The latter is Fourier transformed to the  transverse momentum dependence.
In this mixed representation,
\begin{eqnarray}
  \Sigma_{[ret]}(\tau_1,\tau_2;\eta,\vec{p_t})={ig^{2}\over 2 (2\pi)^2}
  \int d^2 \vec{k_t} [t^{a} \gamma^m(\tau_1) 
  G_{[ret]}(\tau_1,\tau_2;\eta,\vec{p_t}+\vec{k_t})
   t^{b}\gamma^l(\tau_2)
  D^{ba}_{[1]lm}(\tau_2,\tau_1;-\eta,\vec{k_t}) \nonumber \\
 +t^{a} \gamma^m(\tau_1) 
  G_{[1]}(\tau_1,\tau_2;\eta,\vec{p_t}+\vec{k_t})
   t^{b}\gamma^l(\tau_2)
 D^{ba}_{[adv]lm}(\tau_2,\tau_1;-\eta,\vec{k_t})]~,  
\label{eq:E3.4}
\end{eqnarray}
where $\gamma^{\eta}(\tau)=\gamma^3/\tau$.
As it has been shown in [I], all fermion correlators $G_{[\alpha]}$ can be
decomposed as
\begin{eqnarray}
G_{[\alpha]}(\tau_1,\tau_2;\eta,\vec{q_t})=
q_t \big[g^{0}_{[\alpha]}\gamma^0 +g^{3}_{[\alpha]}\gamma^3\big] +
g^{T}_{[\alpha]}q_r\gamma^r + 
i g^{A}_{[\alpha]}q_r\epsilon^{ru}\gamma^u\gamma^5 \nonumber\\
=q_t \big[ g^{L(+)}_{[\alpha]}\gamma^+ 
+g^{L(-)}_{[\alpha]}\gamma^-\big] +
q_r\gamma^r\gamma^0 ~\big[ g^{T(+)}_{[\alpha]}\gamma^+ 
+g^{T(-)}_{[\alpha]}\gamma^-\big]~,
\label{eq:E3.5}
\end{eqnarray}
where, for the sake of brevity, we denote $\vec{q_t}=\vec{p_t}+\vec{k_t}$. 
A similar decomposition takes place for the self-energy,
\begin{eqnarray}
\Sigma_{[ret]}(\tau_1,\tau_2;\eta,\vec{p_t})=
 \Sigma^{0}\gamma^0 +\Sigma^{3}\gamma^3 +
\Sigma^{T}q_r\gamma^r + 
i \Sigma^{A}p_r\epsilon^{ru}\gamma^u\gamma^5 \nonumber\\
= \Sigma^{L(+)}\gamma^+ +\Sigma^{L(-)}\gamma^- +
p_r\gamma^r\gamma^0~\big[ \Sigma^{T(+)}\gamma^+ 
+\Sigma^{T(-)}\gamma^-\big]~,
\label{eq:E3.6}
\end{eqnarray}
and we obviously have,
\begin{eqnarray}
g^{L(\pm)}_{[\alpha]}={1\over 2}(g^{0}_{[\alpha]}\pm g^{3}_{[\alpha]}),~~~~
g^{T(\pm)}_{[\alpha]}={1\over 2}
(g^{T}_{[\alpha]}\pm g^{A}_{[\alpha]})~,~~~~
\Sigma^{L(\pm)}={1\over 2}(\Sigma^{0}\pm \Sigma^{3}),~~~~
\Sigma^{T(\pm)}={1\over 2}(\Sigma^{T}\pm \Sigma^{A})~.
\label{eq:E3.7}
\end{eqnarray}

It becomes easier to analyze the various pieces of the quark self-energy
if the gluon correlators  $D_{[\alpha]lm}$ are taken in the form of the 
following decomposition,\footnote{In what follows, we use the Greek
indices for the four-dimensional vectors and tensors in the curvilinear
coordinates (index $\eta$ is an exception, it always denotes the rapidity
direction), and the Latin indices from $a$ to $d$ for the vectors in flat
Minkowski coordinates. We use Latin indices from $r$ to $w$ for the
transverse $x$- and $y$-components ($r,...,w=1,2$), and the arrows over
the letters to denote the two-dimensional vectors, {\em e.g.}, ${\vec
k}=(k_x,k_y)$, $|{\vec k}|=k_{t}$. The Latin indices from $i$ to $n$
($i,...,n=1,2,3$) will be used  for the three-dimensional internal
coordinates $u^i=(x,y,\eta)$ on the hyper-surface $\tau=const$.}
\begin{eqnarray}
D_{[\alpha]rs}=\bigg(\delta_{rs}- {k_r k_s\over k_t^2}\bigg)
  {\cal D}^{(TE)}_{[\alpha]}+ 
{k_r k_s\over k_t^2}{\cal D}^{(2)}_{[\alpha]},~~~
D_{[\alpha]\eta\eta}={\cal D}^{(\eta\eta)}_{[\alpha]},~~~
D_{[\alpha]r\eta}={k_r\over k_t^2}{\cal D}^{(r\eta)}_{[\alpha]},~~~
D_{[\alpha]\eta s}={k_s\over k_t^2}{\cal D}^{(\eta s)}_{[\alpha]}~,
\label{eq:E3.8}
\end{eqnarray}
where the first term in $D_{[\alpha]rs}$ is due to the transverse 
electric mode, and all invariants of ${\cal D}_{[adv]}$ (except ${\cal
D}^{(TE)}_{[adv]}$ ) have two terms, ${\cal D}_{[0]}^{(\cdot\cdot\cdot)}$
from the  transverse magnetic mode of the radiation field, and ${\cal
D}_{[long]}^{(\cdot\cdot\cdot)}$ from the longitudinal field. All these
components were found in paper [III] and are given in the Appendix in the
form which is used in the calculation below.  After some algebra, we can 
present the retarded quark self-energy in the form, 
\begin{eqnarray}
\Sigma_{[ret]}(\tau_1,\tau_2;\eta,\vec{p_t})={i\alpha_s C_F\over 2\pi} 
\int d^2 \vec{k_t} \big[\gamma^+ S^{L(+)}+\gamma^- S^{L(-)}+ 
 p_r \gamma^r\gamma^0~\big(\gamma^+ S^{T(+)}
 +\gamma^- S^{T(-)}\big)\big]~,   
\label{eq:E3.9} \end{eqnarray} 
where the scalar invariants of $\Sigma_{[ret]}$ are the bilinears of the 
fermion  and gluon scalars,
\begin{eqnarray} 
S^{L(\pm)}=\sum_{[\alpha,\beta]}  \bigg\{q_t
g^{L(\pm)}_{[\alpha]}({\cal D}^{(TE)}_{[\beta]}+{\cal D}^{(2)}_{[\beta]})
+{q_t\over\tau_1\tau_2}g_{[\alpha]}^{L(\mp)} 
{\cal D}^{(\eta\eta)}_{[\beta]} 
\mp{(\vec{k_t} \vec{q_t})\over k_t^2}
\bigg({g^{T(\pm)}_{[\alpha]}{\cal D}^{(r\eta)}_{[\beta]}\over\tau_1}+
{g^{T(\mp)}_{[\alpha]}{\cal D}^{(\eta r)}_{[\beta]}\over\tau_2} 
\bigg)\bigg\}, 
\label{eq:E3.10}\end{eqnarray}
\begin{eqnarray} 
S^{T(\pm)}=\sum_{[\alpha,\beta]}  
\bigg\{\bigg[{(\vec{p_t} \vec{q_t})\over p_t^2} 
-2{(\vec{k_t} \vec{p_t})(\vec{k_t} \vec{q_t})\over k_t^2 p_t^2}\bigg] 
g^{T(\pm)}_{[\alpha]}({\cal D}^{(TE)}_{[\beta]}+{\cal D}^{(2)}_{[\beta]}) 
-{(\vec{q_t} \vec{p_t})\over p_t^2\tau_1\tau_2} 
g^{T(\mp)}_{[\alpha]}{\cal D}^{(\eta\eta)}_{[\beta]}\nonumber\\ 
\mp {(\vec{k_t} \vec{p_t}) \over k_t^2 p_t^2} 
\bigg({g^{L(\pm)}_{[\alpha]} {\cal D}^{(r\eta)}_{[\beta]}\over\tau_1}-
{g^{L(\mp)}{\cal D}^{(\eta r)}_{[\beta]}\over\tau_2}\bigg)\bigg\} ~. 
\label{eq:E3.11}
\end{eqnarray} 
In these equations, the sum $\sum_{[\alpha,\beta]}$ runs over
$[\alpha,\beta]= \{[ret,1], [1, adv]\}$.

\section{Fermion modes in the expanding system }
\label{sec:S2}

We shall look for the dispersion law of the fermions in the proper-time
dynamics studying the Dirac equation (\ref{eq:E3.0b}) with radiative 
corrections. Since the fermions are massless, it is convenient to use the
spinor basis where the Dirac matrices are

\begin{eqnarray}  
\gamma^0= \left( \begin{array}{cc} 0 & I \\ I & 0 \end{array} \right),  
~~~~ \gamma^l= \left( \begin{array}{cc} 0 & -\sigma^l 
                     \\ \sigma^l & 0 \end{array} \right)~, \nonumber 
\end{eqnarray} 
and  the Dirac equation can be split into two 
separate equations for the left- and right-handed two-component spinors.
The latter reads as
\begin{eqnarray}
G_{R}^{-1}(p_t;\tau_1,\eta_1)~\psi_{R}(p_t;\tau_1,\eta_1)=
\int_{0}^{\tau_1}d\tau_2\int_{-\infty}^{\infty}\tau_2 d\eta_2
\Sigma_R(p_t;\tau_1,\tau_2;\eta_1-\eta_2)\psi_{R}(p_t;\tau_2,\eta_2)~,
\label{eq:E2.1}
\end{eqnarray}
where the matrices of the right-handed differential operator $G_{R}^{-1}$ 
and of the right-handed self-energy $\Sigma_{R}$ are
\begin{eqnarray}
G_{R}^{-1}(p_t;\tau,\eta)\left[ \begin{array}{cc}
 i(\partial_\tau+{1\over 2\tau}- 
 {1\over \tau}\partial_\eta) &  p_x-ip_y \\
  p_x+ip_y & i(\partial_\tau+{1\over 2\tau}+ 
  {1\over \tau}\partial_\eta)
           \end{array} \right],\nonumber \\
\Sigma_R(p_t;\tau_1,\tau_2;\eta_1-\eta_2)=
\left[ \begin{array}{cc}
              \Sigma^{L(-)}    & -(p_x-ip_y)\Sigma^{T(+)}  \\
    -(p_x+ip_y)\Sigma^{T(-)}& \Sigma^{L(+)}      \end{array} \right] .	   
\label{eq:E2.2}
\end{eqnarray}
The equation for the left-handed spinors differs from (\ref{eq:E2.1}) only
by  a change of some signs in matrices (\ref{eq:E2.2}) and leads to the
same dispersion law. A solution with positive energy is looked for in the 
form,
\begin{eqnarray}  
\psi_{R}(p_t,\theta;\tau,\eta)=   \left( \begin{array}{c} 
                e^{(\eta-\theta)/2)}~p_t \\
               -e^{-(\eta-\theta)/2)}~(p_x+ip_y) \end{array} \right)
	       ~e^{-i\mu\tau\cosh(\eta-\theta)} ~,
\label{eq:E2.3}\end{eqnarray}
where $\mu$ is the effective ``transverse mass'' of the mode.          
For the free on-mass-shell solution we have $\mu=p_t$.
To solve Eq.~(\ref{eq:E2.1}), we introduce an auxiliary (left-handed) 
spinor
\begin{eqnarray}  
\tilde{\psi}(p_t,\theta';\tau,\eta)=   \left( \begin{array}{c} 
                e^{-(\eta-\theta')/2)}~p_t \\
               -e^{(\eta-\theta')/2)}~(p_x-ip_y) \end{array} \right)
	       ~e^{i\mu\tau\cosh(\eta-\theta')} ~.
\label{eq:E2.4}\end{eqnarray}
We insert (\ref{eq:E2.3}) into (\ref{eq:E2.1}), multiply it from the left
by (\ref{eq:E2.4}) and integrate this along the hypersurface 
$\tau_1=const$.
Then the left side of the equation becomes
\begin{eqnarray} 
\int_{-\infty}^{\infty}\tau_1 d\eta_1
\tilde{\psi}(p_t,\theta`;\tau_1,\eta_1)G_{R}^{-1}(p_t;\tau_1,\eta_1)
\psi_{R}(p_t,\theta;\tau_1,\eta_1)=
4\pi{\mu-p_t\over\mu}~p_t^2~\delta(\theta-\theta')~.
\label{eq:E2.5}\end{eqnarray}
In deriving this equation, we assumed that $\mu$ is independent of $\tau_1$.
The weak dependence is admissible, provided $d\mu/d\tau_1\ll\mu/\tau_1$. 
A solution that has this property does indeed exist. The right hand side 
of the equation is, in fact, independent of $\theta'$ and is of the 
following form,
\begin{eqnarray} 
p_t^2~\int_{0}^{\tau_1}d\tau_2\int_{-\infty}^{\infty}\tau_1 \tau_2 d\eta_2
d\theta d(\eta_1-\eta_2)e^{i\mu\tau_1\cosh(\eta_1+\theta)} 
e^{-i\mu\tau_2\cosh\eta_2} \nonumber \\
\times [e^{-{\eta_1-\eta_2+\theta\over 2}}\Sigma^{(L)-}+
e^{{\eta_1-\eta_2+\theta\over 2}}\Sigma^{(L)+}+
e^{-{\eta_1+\eta_2+\theta\over 2}}\Sigma^{(T)+}+
e^{{\eta_1+\eta_2+\theta\over 2}}\Sigma^{(T)-}]~,
\label{eq:E2.6}\end{eqnarray}
where the exponentials are due to the Thomas precession of the spinor field.
Next, we integrate both sides with respect to $\theta$. Two rapidity
integrals, $d\theta d\eta_2$, on the right absorb the precession factors
yielding the product of Hankel functions,
\begin{eqnarray} 
\pi^2 H^{(1)}_{1/2}(\mu\tau_1)~H^{(2)}_{1/2}(\mu\tau_2)=
{2\pi\over\mu\sqrt{\tau_1\tau_2}}e^{i\mu(\tau_1-\tau_2)}~.
\label{eq:E2.7}\end{eqnarray} 
Finally, we arrive at the dispersion equation that defines the fermion 
``transverse mass'' $\mu$ as a function of the transverse momentum and the
latest time $\tau_1$,
\begin{eqnarray} 
\mu(p_t,\tau_1)-p_t ={1\over 2} 
\int_{0}^{\tau_1}d\tau_2 \sqrt{\tau_1\tau_2} 
e^{i\mu(p_t,\tau_1)(\tau_1-\tau_2)}  \int_{-\infty}^{\infty}d\eta~
[\Sigma^{L(+)}+\Sigma^{L(-)}+p_t \Sigma^{T(+)} +p_t \Sigma^{T(-)}]~.
\label{eq:E2.8}\end{eqnarray}

As it has been discussed in paper [II] for fermions (similar arguments
are true for gluons), only the independence of the quark and gluon
occupation numbers $n_f$ and $n_g$ on rapidity can provide that the
invariants  $S^{L(\pm)}$ and $S^{T(\pm)}$ naturally depend only on the
difference $\eta=\eta_1-\eta_2$. We shall consider only this case of the
local homogeneity;  we can do it safely only because no collinear
singularities which may require a  rapidity cut-off (e.g., $\pm Y$) in
the phase space will appear in the theory. Since we are computing an
essentially local quantity, such a cut-off would be unphysical. With
this reservation, we may rewrite Eq.~(\ref{eq:E2.8}) as
\begin{eqnarray}  
\mu(p_t) = p_t + 
\int_{0}^{\tau_1}d\tau_2 \sqrt{\tau_1\tau_2}  e^{i\mu(p_t)(\tau_1-\tau_2)} 
[\Sigma^{0}(\tau_1,\tau_2)+  p_t \Sigma^{T}(\tau_1,\tau_2)]~,
\label{eq:E3.12}\end{eqnarray} 
where we introduced the notation,
\begin{eqnarray} 
\Sigma^{0}(\tau_1,\tau_2)=
{i\alpha_s C_F\over 4\pi}\sum_{[\alpha,\beta]} \int d^2 \vec{k_t}
\int_{-\infty}^{\infty}d\eta~ q_t g^{0}_{[\alpha]}
[{\cal D}^{(TE)}_{[\beta]}+{\cal D}^{(2)}_{[\beta]} 
+{1\over \tau_1\tau_2}{\cal D}^{(\eta\eta)}_{[\beta]}],
\label{eq:E3.13}\end{eqnarray}
\begin{eqnarray} 
\Sigma^{T}(\tau_1,\tau_2)={i\alpha_s C_F\over 4\pi}\sum_{[\alpha,\beta]} 
\int d^2\vec{k_t}\int_{-\infty}^{\infty}d\eta~
g^{T}_{[\alpha]} \bigg\{
\bigg[{(\vec{p_t} \vec{q_t})\over p_t^2} 
-2{(\vec{k_t} \vec{p_t})(\vec{k_t} \vec{q_t})\over k_t^2 p_t^2}\bigg]
({\cal D}^{(TE)}_{[\beta]} +{\cal D}^{(2)}_{[\beta]})+ 
{(\vec{q_t} \vec{p_t})\over p_t^2\tau_1\tau_2}
{\cal D}^{(\eta\eta)}_{[\beta]}\bigg\}.
\label{eq:E3.13a}\end{eqnarray}
Comparing these equations with Eqs.~(\ref{eq:E3.10}) and (\ref{eq:E3.11}), we
may observe a significant simplification. The terms with the off-diagonal
components ${\cal D}^{(\eta r)}$ and  ${\cal D}^{(r\eta)}$  have dropped out.
These terms, as it can be seen from Eqs.~(\ref{eq:A1.6})-- (\ref{eq:A1.7}),
(\ref{eq:A1.10})--(\ref{eq:A1.11}), and 
(\ref{eq:A1.16})--(\ref{eq:A1.17}), are odd with respect to $\eta$, while
the invariants $g^{0}=g^{L(+)}+g^{L(-)}$ and  $g^{T}=g^{T(+)}+g^{T(-)}$ are
even. Therefore, integration over $\eta$ eliminates the terms with the
off-diagonal components.

\section{Propagators, densities of states, and occupation numbers
         in the expanding system }
\label{sec:S4}

In this section, we collect condensed information about various correlators
of quark and gluon fields derived in papers [II] and [III] which are
necessary for the calculation of the quark self-energy. We also discuss our
specific choice of  occupation numbers $n_g(k_t,\alpha)$ and
$n_f(q_t,\theta)$.  All field correlators are defined as the expectation
values over the distribution of the background particles. The latter are
the excitations of the modes allowed by the constraints and the boundary
conditions of wedge dynamics. The Fock space of these excitations was
constructed in papers [II] and [III]. We have analyzed two sets of quantum
numbers that may label the states. Both sets include the transverse
momentum $\vec{p_t}$ and polarization index. In one set, the remaining
variable was the boost $\nu$ (the variable conjugated to the coordinate
$\eta$); this set proved to be very useful in the practical calculation of
the gluon propagators.  In the second set, the particles are labeled by
their velocity $v_z=\tanh\theta$ in the direction of the collision axis.
This representation is used below. The fermion spectral functions are
\begin{eqnarray}
 G_{[10]}(q_t,\theta;\tau_1,\tau_2) =[1-n_f^+(q_t,\theta)]
 G^{(0)}_{[10]}(q_t,\theta;\tau_1,\tau_2) -
 n_f^-(q_t,\theta) G^{(0)}_{[01]}(q_t,\theta;\tau_1,\tau_2)~~,\nonumber\\
 G_{[01]}(q_t,\theta;\tau_1,\tau_2) = -n_f^+(q_t,\theta)
 G^{(0)}_{[10]}(q_t,\theta;\tau_1,\tau_2) + [1-n_f^-(q_t,\theta)]
 G^{(0)}_{[01]}(q_t,\theta;\tau_1,\tau_2)~.
\label{eq:E4.1}\end{eqnarray}
Their gluon counterparts are of a similar form,
\begin{eqnarray}
D_{[10]}(k_t,\alpha;\tau_1,\tau_2) =[1+n_g(k_t,\alpha)]
 D^{(0)}_{[10]}(k_t,\alpha;\tau_1,\tau_2) +
 n_g(q_t,\alpha) D^{(0)}_{[01]}(q_t,\alpha;\tau_1,\tau_2)~,\nonumber\\
D_{[01]}(k_t,\alpha;\tau_1,\tau_2) =n_g(k_t,\alpha)
 D^{(0)}_{[10]}(k_t,\alpha;\tau_1,\tau_2) +
[1+ n_g(q_t,\alpha)] D^{(0)}_{[01]}(q_t,\alpha;\tau_1,\tau_2)~, 
\label{eq:E4.2}
\end{eqnarray}
where $D^{(0)}_{[\alpha]}$ and $G^{(0)}_{[\alpha]}$ are the vacuum 
correlators of a given type $[\alpha]$.  They are defined as  vacuum 
expectation values  of the binary products of field operators,
\begin{eqnarray}
G^{(0)}_{[10]}(x_1,x_2) = 
-i\langle 0|\Psi (x_1)\overline{\Psi}(x_2)|0\rangle~,~~~~~
G^{(0)}_{[01]}(x_1,x_2)=
i\langle 0|\overline{\Psi}(x_2)\Psi (x_1)|0\rangle~, \nonumber\\
D^{(0)}_{[10]lm}(x_1,x_2) = 
-i\langle 0|A_ll(x_1) A(x_2)_m|0\rangle~,~~~~
D^{(0)}_{[01]lm}(x_1,x_2)=-i\langle 0|A_m(x_2)A_l(x_1)|0\rangle~.
\label{eq:E4.2a}
\end{eqnarray}
In this approximation, the field (anti-)commutators,
\begin{eqnarray}
 G_{[0]}= G_{[10]}-G_{[01]}=G^{(0)}_{[10]}-G^{(0)}_{[01]}=
 G^{(0)}_{[0]}~,\nonumber\\
 D_{[0]}= D_{[10]}-D_{[01]}=D^{(0)}_{[10]}-D^{(0)}_{[01]}=
 D^{(0)}_{[0]}~
\label{eq:E4.3}
\end{eqnarray}
appear to be insensitive to the presence of the particle distribution,
while their counterparts,
\begin{eqnarray}
 G_{[1]}= G_{[10]}+G_{[01]}= [1-2n_f]G^{(0)}_{[1]}~,\nonumber\\
 D_{[1]}= D_{[10]}+D_{[01]}= [1+2n_g]D^{(0)}_{[1]}~,
\label{eq:E4.4}
\end{eqnarray}
include the occupation numbers which modify the original vacuum density
of states. For the sake of simplicity, we take $n_f^+=n_f^-=n_f$, which 
corresponds to a neutral system. 

The Wightman functions (\ref{eq:E4.1}) and (\ref{eq:E4.2}) (or their
various linear combinations $G_{[\beta]}$ and $D_{[\beta]}$) eventually
appear under the integrals $d\theta$ and $d\alpha$. One must keep in mind 
that  in order to reduce  $G^{(0)}_{[\beta]}$ and $D^{(0)}_{[\beta]}$ to
the standard form of the vacuum correlators, at least two shifts of the
integration variables is necessary. Only after that will
$G^{(0)}_{[\beta]}$ and $D^{(0)}_{[\beta]}$  explicitly depend on the
boost-invariant variables $\eta$ and $\tau_{12}$.  The functions
$n_g(k_t,\alpha)$ and $n_f(q_t,\theta)$ are not indifferent to this shift.
It may well happen that a formal shift in $\theta$ or $\alpha$ will drive
the stationary points of the wave functions or the singularities of the
field correlators outside the physical boundaries of the distributions
$n_g(k_t,\alpha)$ and $n_f(q_t,\theta)$. Therefore, different
representations of $G_{[1]}$ and $D_{[1]}$ must be used for the study of
different subprocesses. One has to account for the reservations stemming
from the derivation procedure described in Sec.~4 of  paper [II]. These
different representations of the quark and gluon correlators are quoted in
the Appendix.

In our picture, first outlined in paper [I],  the fermion vacuum mode with
small transverse momentum $p_t$ and zero rapidity is modified by its
forward scattering either on gluons with high momentum $k_t$ and rapidity
$\alpha$, $k_t\gg p_t$ or on quarks with  high momentum $q_t$ and rapidity
$\theta$, $q_t\gg p_t$.  These hard modes are created at the earliest
moment of the collision and can be treated as well formed particles by the
time $\tau\sim 1/p_t$, since at that time $\tau k_t\gg 1$, and $\tau q_t\gg
1$. Therefore, they may be consistently described by the distributions,
\begin{eqnarray}
n_f(q_t,\theta)\approx {{\cal N}_f\over \pi R_\bot^2}
{\theta (q_t-p_\ast) \over q_t^2},~~~
 n_g(k_t,\alpha)\approx {{\cal N}_g\over \pi R_\bot^2}
 {\theta (k_t-p_\ast) \over k_t^2}~.
\label{eq:E4.5}
\end{eqnarray}
where $p_\ast$ is the lower bound of the ``hard'' partons distribution.
Both distributions (per unit area, per unit rapidity) are chosen on
purely dimensional grounds, since we believe that the creation of a parton
with large transverse momentum is described by perturbative QCD which has 
no intrinsic scale. 

Currently, the normalization factors ${\cal N}_g$ and ${\cal N}_f$ are the
only (apart from the coupling $\alpha_s$) parameters of the theory. The
cross section $\pi R_\bot^2$ and the full width  $2Y$ of the rapidity
plateau are defined by the geometry of a particular collision and the
c.m.s. energy, respectively. These  are irrelevant for the local screening
parameters we are interested in. In the first approximation, one may try to
extract them from the event-by-event measurement of the high-$p_t$ tail of
the collision products and incorporating the standard phenomenology of the
fragmentation functions for the analysis.

As it was pointed out in paper [I], even in dense systems, the QCD
evolution at large $Q^2$ is not likely to be affected by finite-density
effects. Thus, one may also try to employ the known structure functions
(without shadowing corrections) and the factorization scheme in order to
estimate  ${\cal N}_g$  and ${\cal N}_f$. A most appealing opportunity to
find $n_g(k_t,\alpha)$ and $n_f(q_t,\theta)$ from first principles,
associating them with the known properties of hadrons and the QCD vacuum,
is still very distant. 

The distributions   (\ref{eq:E4.5})  are used below with the following
informal reservations. First, the total energy of any  collision is finite
and $k_t$  and $q_t$ have (though very high, but finite) upper boundary.
Eventually, this leads to the self-energy which is free from collinear
singularities in the interaction of charges with the vector gauge field.
Second, though the distributions (\ref{eq:E4.5}) are boost-invariant, only
the particles which physically affect the forward scattering  must be
accounted for. There is  a strong correlation between the position $\eta$
where the particle with large transverse momentum $q_t$ is measured (or is
interacting) and its rapidity $\theta$. Hence, the limits of integrals
$d\alpha$ and $d\theta$ over the rapidities of real particles (which either
mediate the scattering or are in the final states) cannot exceed the actual
rapidity boundaries of the scattering process.  In its turn, this puts an
additional requirement on the notion of the distribution itself. It must be
normalizeable in the physical volume of the reaction. This volume is
defined, in fact, by the light cone (i.e. causality of the forward
scattering amplitude). [We remind the reader that the notion of a
distribution itself makes sense only after it is prepared (measured) at
least in a {\em gedanken} experiment. Hence, the distributions  $n_g$ and
$n_f$ must exist, in this sense, both at final time $\tau_1$ and at the
initial time $\tau_2$ in the expression for the self-energy. In its turn,
this limits the time $\tau_2$ from below.]

\section{Leading part of the dispersion equation}
\label{sec:S3a}

\subsection{Derivation of the dispersion equation}
\label{subsec:Sb3a}

\begin{figure}[htb]
\begin{center}
\mbox{ \psfig{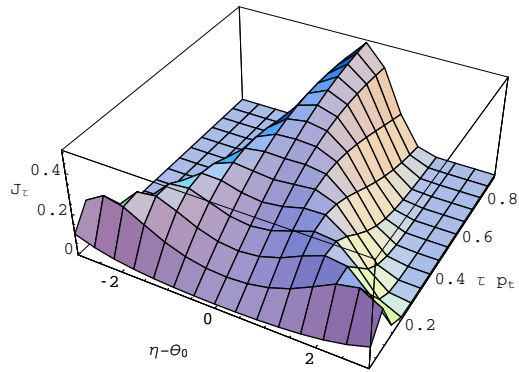} }
\end{center}
\caption{Evolution of the charge density in the typical state of wedge
dynamics}
\label{fig:fig0}\end{figure}

The most important outcome of this work is that the major contribution to
the effective quark mass comes from the $\eta\eta$-component of the
propagator of the longitudinal field. This contribution is computed in all
details below. All other terms are associated with the propagation of the
transverse fields and they appear to be parametrically small in the domain
$\tau_1p_t <1$, $(\tau_1-\tau_2) p_t \ll 1$, $\tau_1-\tau_2 \ll\tau_1$,
where the dynamical mass of the fermion is effectively formed.  The
component $D_{\eta\eta}$ of the propagator establishes the connection
between the $A_\eta$ component of the potential and the $j_\eta$ component
of the current. In its turn, $A_\eta$ is responsible for the
$\eta$-component $E_\eta=\partial_\tau A_\eta$ of the electric field and
the $x$- and $y$-components, $B_x=\partial_y A_\eta$,  $B_y=-\partial_x
A_\eta$ of the magnetic field. The electrical field in the longitudinal
$\eta$-direction is not capable of producing scattering with transverse
momentum transfer. However, this transfer can be provided by the magnetic
forces; the two currents $j_\eta$ can interact via the magnetic field
${\vec B}_t=(B_x,B_y)$. The origin of these currents is intrinsically
connected with the geometry of states in the wedge form of dynamics. Any
state with a given $p_t$ begins its life being widely spread along the
light cone. If the state is charged, then  local charge density is small.
With time going on, the spread of the wave function diminishes and the
charge become localized in a narrower rapidity interval (see
Fig.~\ref{fig:fig0}). Therefore, any  charged state carries a current in
the longitudinal (rapidity) direction.  

The magnetic fields of the {\em transition currents} provide scattering
with  the most effective transfer of the transverse momentum. Indeed, at
time $\tau_2$ a quark with the transverse momentum $p_t$, $\tau_2 p_t\ll
1$, interacts with the gluon field and acquires a large transverse momentum
$k_t$, $\tau_2 k_t\gg 1$. This transition is characterized by a drastic
narrowing of the charge spread in the rapidity direction, and must be
accompanied by a strong $\eta$-component of the transition current. A
similar transition in the opposite direction happens at time $\tau_1$, when
the gluon field interacts with another quark that has large initial
transverse momentum $k_t$, and recovers the soft state with $\tau_2 p_t\ll
1$ in the course of this interaction. This second transition current
readily interacts with the magnetic component of the gluon field. Our
estimates indicate that the leading contribution comes  from the term of
$D_{\eta\eta}(\tau_2,\tau_1; \eta_2-\eta_1;\vec{k_t})$   which is
proportional to $\delta (\eta_1-\eta_2)$ and does not depend on  ${\vec
k_t}$ [in coordinate  representation, this term is just  proportional to
$\delta (\eta_1-\eta_2) \delta (\vec{r_1}-\vec{r_2})$ ].  This is a
long-range contact interaction of the two currents, and is  not limited by
the light-cone boundaries (which suppress the interaction  via the strongly
localized states of the radiation field). Furthermore, the  contact part of
the longitudinal propagator is the only one that brings  into the integrand
of the dispersion equation (\ref{eq:E3.12}) the term  which is {\em
singular} at $\tau_1-\tau_2\to 0$. Therefore, it is capable  of providing
an appreciable contribution into the effective quark mass, which is {\em
defined locally}. This  part of the self-energy allows for an exact
calculation with a simple  analytic answer which is presented below.
Estimates of all small terms  are  explained in Sec.~\ref{sec:S6} and
\ref{sec:SA2}\footnote{ The authors appreciate discussions with Edward
Shuryak, who pointed out that the small effect of the radiation fields is
much less surprising than the finite contribution from the longitudinal
fields.}. This contact part  of the longitudinal propagator is
\begin{eqnarray} 
D^{[contact]}_{\eta\eta}(\tau_2,\tau_1;\eta_2-\eta_1;\vec{k_t})
=-{\tau_1^2-\tau_2^2 \over 2}\delta (\eta).
\label{eq:E3.14}\end{eqnarray}
Because of the extreme locality of $D^{[contact]}_{\eta\eta}$ provided 
by $\delta(\eta)$, the invariants $g_{[1]}$ of the fermion density function
in (\ref{eq:E3.13}) lose their kinematic coefficients, 
\begin{eqnarray} 
g^{0}_{[1]}= {-2{\cal N}_f\over \pi R_t^2}~
{Y_1(\tau_{12}q_t)\over q_t^2}~,~~~~~~
g^{T}_{[1]}= {-2{\cal N}_f\over \pi R_t^2}~{Y_0(\tau_{12}q_t)\over q_t^2}~.
\label{eq:E3.16}\end{eqnarray}
First, we integrate over $\eta$, which leads to $\tau_{12}^{2}=
(\tau_1-\tau_2)^2$. Next, we change $d^2\vec{k_t}$ for $d^2\vec{q_t}$ and
integrate over the orientation of $\vec{q_t}$ gaining the factor $2\pi$ in
$\Sigma^{0}$. In $\Sigma^{T}$, $g^{T}_{[1]}$ is integrated with the weight
factor $(\vec{q}_t\cdot\vec{p}_t)/p_t^2$. Therefore, this term identically 
vanishes after integration over the azimuthal angle. The only remaining 
integral over the transverse momenta of hard partons is
\begin{eqnarray} 
\int_{p_\ast}^{\infty} Y_1[(\tau_1 -\tau_2)q_t]~d q_t =
{Y_0[(\tau_1 -\tau_2)p_\ast]\over\tau_1 -\tau_2}~.
\label{eq:E3.17}\end{eqnarray}
Eventually, we may write the dispersion equation (\ref{eq:E3.12}) as
follows,
\begin{eqnarray} 
\mu = p_t +  {i\alpha_s C_F {\cal N}_f\over 2\pi R_t^2}
\int_{0}^{\tau_1}d\tau_2 {\tau_1+\tau_2\over 2\sqrt{\tau_1\tau_2}} 
e^{i\mu(\tau_1-\tau_2)} Y_0[(\tau_1 -\tau_2)p_\ast] ~.
\label{eq:E3.18}\end{eqnarray}

\subsection{Study of the dispersion equation.}
\label{subsec:Sb3b} 

According to the qualitative analysis of paper [II], the dynamics of states
is different in the two limiting cases, $\tau p_t <1$ and $\tau p_t >1$.
With respect to $p_t\sim 1/\tau$, the states are divided into ``hard'' and
``soft'' states. Therefore, it is natural to take $p_\ast\sim p_t$ in
Eq.~(\ref{eq:E3.18}). Further, it is convenient to trade variable $\tau_2$ 
for $y=(\tau_1 -\tau_2)/\tau_1$,
\begin{eqnarray} 
{\mu(p_t,\tau_1)\over p_t} = 1+ 
{i\alpha_s C_F {\cal N}_f \tau_1 p_t\over 2\pi R_t^2 p_t^2}
\int_{0}^{1}dy {1-y/2\over \sqrt{1-y}}
e^{i\mu(p_t,\tau_1)\tau_1y} Y_0(\tau_1 p_t y) ~.
\label{eq:E3.19}\end{eqnarray}
In this form, the dispersion equation clearly reveals two  distinctive
regimes. When $\tau p_t <1$, then the function $Y_0(x)$ behaves as a
logarithm, and the right hand side of Eq.~(\ref{eq:E3.19}) becomes
proportional to $\ln(2/\tau_1 p_t)$, the effective width of the rapidity
interval occupied by the state at the early time of the evolution. When
$\tau p_t >1$, then $Y_0(x)\sim 1/\sqrt{x}$, and the integral becomes
proportional to $1/\sqrt{\tau p_t}$, the effective rapidity width at
later times. Thus, the dispersion equation (\ref{eq:E3.19}) clearly
reveals  two distinctive regimes which were qualitatively analyzed in paper
[II].  The solution of Eq.~(\ref{eq:E3.19}) is generally complex. Taking
$\mu=\mu'+i\mu''$ we can separate real and imaginary parts of this 
equation,
\begin{eqnarray} 
\tau_1\mu'- \tau_1 p_t = 
- {\alpha_s C_F {\cal N}_f  \over 2\pi (R_t^2 /\tau_1^2)}
\int_{0}^{1}{dy\over 2}~\bigg[ ~{1\over \sqrt{1-y}}+\sqrt{1-y}\bigg] ~
e^{-\mu''\tau_1 y}~
\sin (\mu'\tau_1 y) Y_0(\tau_1 p_t y) ~,
\label{eq:E3.20}\end{eqnarray}
\begin{eqnarray}
\tau_1\mu''  =  
{\alpha_s C_F {\cal N}_f \over 2\pi (R_t^2/\tau_1^2)}
\int_{0}^{1}dy ~\bigg[ ~{1\over \sqrt{1-y}}+\sqrt{1-y}\bigg]~
e^{-\mu'' \tau_1 y}~
\cos (\mu'\tau_1 y) Y_0(\tau_1 p_t y) ~,
\label{eq:E3.21}
\end{eqnarray}
(The unit upper limit in these integrals corresponds to $\tau_2=0$, and is,
as a matter of fact,  fictitious. Practically, we are interested only in 
the domain where $\tau_2 p_t\sim 1$.)  We have rearranged the factor  in
front of the integral in such a way, that at early times, this factor is
small. It has been shown in paper [III], that the longitudinal part of the
gluon propagator vanishes when the distance $r_t$ exceeds $\tau_1$.
Therefore, this factor is proportional to the  (small) number of hard
partons per transverse area occupied by the soft quark  mode. Hence, we can
analyze Eqs,~(\ref{eq:E3.20}) and (\ref{eq:E3.21}) by successive
approximations. It is clear, that in the lowest approximation, we can take
$\mu'=p_t$ in the RHS of these equations, and that the imaginary part
$\mu''$ can be neglected. Using  $$\sin x\approx x~,~~~ \cos x\approx 1~,
~~~ Y_0(x) \approx 2\pi^{-1} [\gamma_E +\ln x]~, $$ as an approximation,
and computing the remaining integrals, we arrive at
\begin{eqnarray} 
{\mu'\over p_t} -1 = 
 {\alpha_s C_F {\cal N}_f \over \pi^2 R_t^2 p_t^2}~(\tau_1 p_t)^2~ 
\bigg[ {4\over 5}\bigg( -\gamma_E +\ln{2\over \tau_1 p_t}\bigg) 
+{104-120\ln 2\over 75}\bigg]~,
\label{eq:E3.20a}\end{eqnarray}
\begin{eqnarray}
{\mu'' \over p_t} =  
{\alpha_s C_F {\cal N}_f \over \pi^2 R_t^2 p_t^2}~\tau_1 p_t~
\bigg[{4\over 3}(\gamma_E -\ln{2\over \tau_1 p_t}\bigg) 
+ {26 - 24\ln 2\over 9}\bigg]~.
\label{eq:E3.21a}
\end{eqnarray}
These dependences are plotted below as  functions
of the argument $\tau_1 p_t$ up to the pre-factor
$\alpha_s C_F {\cal N}_f /\pi^2 R_t^2 p_t^2 $.

\begin{figure}[htb]
\begin{center}
\mbox{ 
\psfig{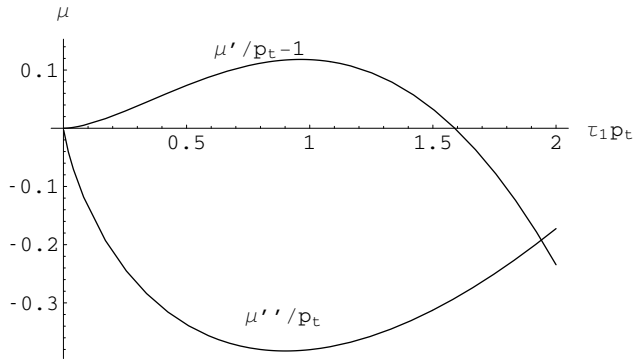}
}
\end{center}
\caption{Self-energy corrections to the real part ($\mu'/ p_t -1$,
upper curve) and to the imaginary part ($\mu''/ p_t $,
lower curve) of the effective transverse mass as  functions
of $\tau_1p_t$.}
\label{fig:fig1}
\end{figure}
Eq.~(\ref{eq:E3.19}) describes the evolution of the effective transverse
mass $\mu$ of the state with a given transverse momentum $p_t$ as a
function of the proper time $\tau_1$. We see, that the real part $\mu'$
gradually grows with time reaching its maximum at $\tau_1p_t\approx 1$. The
mode acquires an ``adjoint mass'' due to the interaction with  hard
partons, as  was anticipated. The curves cannot be trusted above the
boundary $\tau_1\approx p_t^{-1}$, since at  later times, the mode becomes
``hard''. It cannot be viewed as a soft cloud swept with  uniformly
distributed hard particles. The condition $$ {d\mu \over d\tau_1}\ll{\mu
\over \tau_1},$$ which was used in the course of the dispersion equation
derivation, is clearly fulfilled near the maximum of the dispersion curve.

One more important dependence is hidden in the pre-factor 
$\alpha_s C_F {\cal N}_f /\pi^2 R_t^2 p_t^2 $, and is not visible from the
figure above. This factor scales as $p_t^{-2} $, clearly indicating that at
large $p_t$, the effect of screening is small. There is almost no hard
particles with $k_t,q_t>p_t$.

\section{Cancelation of collinear terms in the vacuum part of 
          the self-energy}
\label{sec:S5}

Usually, the self-energy $\Sigma$ is studied in the momentum
representation, and the first subject of concern is the ultraviolet
divergence of this function. It is well known that this divergence can be
at most logarithmic. Thus, when we compute the four-dimensional integral
over the momentum in the loop, this divergence can show up only in the last
of these integrations. We compute the self-energy in the mixed
representation. Hence, we cannot see the UV divergence explicitly but we
{\em must} already see (e.g., in ${\rm Im}\Sigma$) the various infrared
divergences which emerge due to real processes with massless fields. A
corresponding analysis for the case of the null-plane dynamics  was done in
paper [I]. These divergences must be regularized (or even removed from the
theory, as is done by means of dimensional regularization) before the UV
renormalization. The primary goal of this section is to demonstrate that in
the wedge form of dynamics the quark self-energy is completely protected
from  collinear problems, and that this is not a surprise. Indeed,  in the
theory with massless fermions and gauge bosons, the infrared singularities
show up in a different way depending on the type of Hamiltonian dynamics
(including the gauge condition) which is used to describe the process. In
the gauge $A^+ =0$ they look like collinear divergences. In the gauge $A^0
=0$, they look like an infrared problem of the proper field of the charged
particle. In both cases, the problem emerges due to the incomplete gauge
fixing, and manifests itself through  spurious poles of the gauge field
propagators. As  has been shown in paper [III], the gauge $A^\tau =0$ is
fixed completely, and therefore, the quark self-energy that we compute here
is totally free of these problems.

In order to demonstrate this appealing feature we shall compute (the most
dangerous in this respect) the vacuum part of the fermion self-energy,
concentrating on the terms where the integrand as a function of the
rapidity $\alpha$ is not suppressed at $~|\alpha|\to\infty~$. A
self-consistent piece of this type is the contribution of the transverse
electric mode of the gluon field. The tensor part of any gluon correlator
for this mode is of a very simple form,
\begin{eqnarray}
 D_{[\alpha]rs}^{TE}=(\delta_{rs} -k_rk_s/k_t^2)~{\cal D}_{[\alpha]}^{TE}~;
\label{eq:E5.1}
\end{eqnarray}
it has no $\eta$ components, and the scalar functions
${\cal D}_{[\alpha]}^{TE}$ can be computed exactly since they 
have simple integral representations. 
We use this piece of the self-energy to explain the principles we base our
calculations on.
Using Eqs.~(\ref{eq:E3.13}) and (\ref{eq:E3.13a}), we get
\begin{eqnarray}
\big[\Sigma_{[ret]}^{L(\pm)}\big]^{TE}_{vac}(\tau_1,\tau_2;\eta,\vec{p_t})=
{i\alpha_sC_F\over 2\pi}\theta(\tau_{12}^2)\theta(\tau_1-\tau_2)
\int d^2\vec{k_t}\big[q_t g^{L(\pm)}_{[0]}(q_t){\cal D}_{[1]}^{TE}(k_t)-
q_t g^{L(\pm)}_{[1]}(q_t){\cal D}_{[0]}^{TE}(k_t)\big]~,
\label{eq:E5.3}
\end{eqnarray}
where $\vec{q_t}=\vec{k_t}+\vec{p_t}$ and the minus sign in the second term
is due to the definition (\ref{eq:E3.2}) of $D_{[adv]}$. The 
$\Sigma_{[ret]}^{TE}$ is fully confined within the light wedge 
$\tau_{12}^2 >0$.
Then, the vacuum quark and gluon correlators have the following form,
\begin{eqnarray}
{\cal D}_{[0]}^{TE}(\tau_2,\tau_1;\eta_2-\eta_1;\vec{k_t})=
2^{-1}\theta(\tau_{12}^{2})J_0(\tau_{12}k_t)~,~~~~
{\cal D}_{[1]}^{TE}(\tau_2,\tau_1;\eta_2-\eta_1;\vec{k_t})=
2^{-1}i~Y_0(\tau_{12}k_t)~,
\label{eq:E5.2a}
\end{eqnarray}
\begin{eqnarray}
 g^{L(\pm)}_{[0]}=
 i~{\tau_1 e^{\mp\eta/2}-\tau_2e^{\pm\eta/2}\over 4\sqrt{|\tau_{12}^2|}}
 \theta(\tau_{12}^2)~J_1(q_t\sqrt{|\tau_{12}^2|})~,~~~~
 g^{L(\pm)}_{[1]}= 
  {\tau_1 e^{\mp\eta/2}-\tau_2e^{\pm\eta/2}\over 4\sqrt{|\tau_{12}^2|}}
  ~Y_1(q_t\sqrt{|\tau_{12}^2|})~,
\label{eq:E5.2b}
\end{eqnarray}
where $\tau_{12}^2=\tau_{1}^2+\tau_{2}^2 - 2\tau_1\tau_2\cosh\eta$, and
the commutators $D_{[0]}^{TE}$ and $G_{[0]}$ include the causal  
$\theta(\tau_{12}^{2})$ by  definition, and the terms proportional
to $\theta(-\tau_{12}^{2})$ in the densities $D_{[1]}^{TE}$ and
$G_{[1]}$ are omitted.
Integration over the angle $\varphi$ between the vectors $\vec{p_t}$ and 
$\vec{k_t}$ involves only the invariants $g_{[\alpha]}$. According to
Eqs.(\ref{eq:E5.3}) and (\ref{eq:E5.2b}), we have to integrate,
\begin{eqnarray}
\int_{0}^{2\pi}q_t J_1(\tau_{12}q_t)d\varphi=
2\pi~[ k_t J_0(\tau_{12}p_t)J_1(\tau_{12}k_t)+
p_t J_1(\tau_{12}p_t)J_0(\tau_{12}k_t)]~,\nonumber\\
\int_{0}^{2\pi}q_t Y_1(\tau_{12}q_t)d\varphi=
2\pi~\{\theta(k_t-p_t)[ k_t J_0(\tau_{12}p_t)Y_1(\tau_{12}k_t)+
p_t J_1(\tau_{12}p_t)Y_0(\tau_{12}k_t)]+\nonumber\\
+\theta(p_t-k_t)[ k_t J_1(\tau_{12}k_t)Y_0(\tau_{12}p_t)+
p_t Y_1(\tau_{12}p_t)J_0(\tau_{12}k_t)] \}~.
\label{eq:E5.4}
\end{eqnarray}
This integration is done with the aid of the so called addition 
theorems \cite{Watson} for Bessel functions of the argument
$~q_t=[k_t^2+p_t^2+2k_tp_t\cos\varphi]^{1/2}$. Starting from this point, we
can continue in two ways. The most straightforward option is to use the
gluon correlators in the integrated form of the Bessel functions, thus 
sweeping under the rug the singular behavior stemming from 
$\alpha\to\infty$. This leads to the integral
\begin{eqnarray}
\big[\Sigma_{[ret]}^{L(\pm)}\big]^{TE}=
{i\alpha_sC_F\over 8}\theta(\tau_{12}^2)\theta(\tau_1-\tau_2)
{\tau_1 e^{\mp\eta/2}-\tau_2e^{\pm\eta/2}\over \tau_{12}}\nonumber\\
\times\bigg[ - J_0(\tau_{12}p_t)
\int_{0}^{\infty}k_t^2[J_1(\tau_{12}k_t)Y_0(\tau_{12}k_t)+
J_0(\tau_{12}k_t)Y_1(\tau_{12}k_t)]dk_t -
2p_tJ_1(\tau_{12}p_t)
\int_{0}^{\infty}k_tJ_0(\tau_{12}k_t)Y_0(\tau_{12}k_t)dk_t\nonumber\\
+p_tJ_1(\tau_{12}p_t)
\int_{0}^{p_t}k_tJ_0(\tau_{12}k_t)Y_0(\tau_{12}k_t)dk_t+
J_0(\tau_{12}p_t)
\int_{0}^{p_t}k_t^2J_0(\tau_{12}k_t)Y_1(\tau_{12}k_t)dk_t\nonumber\\
-p_t Y_1(\tau_{12}p_t)
\int_{0}^{p_t}k_tJ_0(\tau_{12}k_t)J_0(\tau_{12}k_t)dk_t
-Y_0(\tau_{12}p_t)
\int_{0}^{p_t}k_t^2J_0(\tau_{12}k_t)J_1(\tau_{12}k_t)dk_t\bigg]
 \label{eq:E5.5}
\end{eqnarray}
Here, all the integrals can be computed explicitly as indefinite integrals.
The integrals from $0$ to $p_t$ yield the regular part of the answer
below. Taking the upper limit of improper integrals to be $\Lambda\to\infty$,
we get the singular part as the limit,
\begin{eqnarray}
\lim_{\Lambda\to\infty} {\Lambda^2\over\tau}\bigg[[J_0(\tau_{12}p_t)
+\tau p_tJ_1(\tau_{12}p_t)]J_1(\tau\Lambda)Y_1(\tau\Lambda)+
\tau p_tJ_1(\tau_{12}p_t)J_0(\tau\Lambda)Y_0(\tau\Lambda)\bigg]~.
\label{eq:E5.6}
\end{eqnarray}
Using the asymptotic expansion of Bessel functions, we find that
the singular part is built from $\delta(\tau_{12})$ and its derivative.
The full answer reads as
\begin{eqnarray}
\big[\Sigma_{[ret]}^{L(\pm)}\big]^{TE}=
-{i\alpha_sC_F\over 16\pi}\theta(\tau_{12}^2)\theta(\tau_1-\tau_2)
{\tau_1 e^{\mp\eta/2}-\tau_2e^{\pm\eta/2}\over \tau_{12}}\nonumber\\
\times\bigg\{ p_t^2 {J_2(\tau_{12}p_t)\over\tau_{12}}
+\pi\bigg[{J_0(\tau_{12}p_t)\over\tau_{12}}\bigg(\delta'(\tau_{12})
-{2\over\tau_{12}}\delta(\tau_{12})\bigg)-
{4p_tJ_1(\tau_{12}p_t)\over\tau_{12}}\delta(\tau_{12})\bigg]\bigg\}~.
\label{eq:E5.7}
\end{eqnarray}
Thus, the answer is singular at the null planes $\tau_{12}^2=0$. In order to
find the true source of this singularity, we shall proceed in a different
manner, keeping the gluon invariants $g^{L(\pm)}$ in the integral form. We
shall integrate over $k_t$ first and leave the integral over the gluon
rapidity $\alpha$ for the end of calculation. This leads to
\begin{eqnarray}
\big[\Sigma_{[ret]}^{L(\pm)}\big]^{TE}=
{i\alpha_sC_F\over 8}\theta(\tau_{12}^2)\theta(\tau_1-\tau_2)
{\tau_1 e^{\mp\eta/2}-\tau_2e^{\pm\eta/2}\over \tau_{12}}
\bigg\{-~{ p_t^2\over 2\pi} {J_2(\tau_{12}p_t)\over\tau_{12}}\nonumber\\
+\int_{-\infty}^{\infty}{d\alpha\over\pi}~
\bigg[ J_0(\tau_{12}p_t)
\bigg(\int_{0}^{\infty}k_t^2J_1(\tau_{12}k_t)\cos (T_{12}k_t) dk_t-
\int_{0}^{\infty}k_t^2 Y_1(\tau_{12}k_t)
\sin (T_{12}k_t) dk_t\bigg)\nonumber\\
+p_tJ_1(\tau_{12}p_t)
\bigg(\int_{0}^{\infty}k_tJ_0(\tau_{12}k_t)\cos (T_{12}k_t) dk_t-
\int_{0}^{\infty}k_t Y_0(\tau_{12}k_t)
\sin (T_{12}k_t) dk_t\bigg)\bigg]\bigg\}~,
 \label{eq:E5.8}
\end{eqnarray}
where, we remind the reader that, 
$T_{12}=T_1-T_2=\tau_{12}\cosh(\alpha-\psi)$. The integrals
in this expression are the well-known Fourier transforms of the Bessel
functions,
\begin{eqnarray}
\int_{0}^{\infty}k_t^2J_1(\tau_{12}k_t)\cos (T_{12}k_t) dk_t=
\int_{0}^{\infty}k_t^2 Y_1(\tau_{12}k_t)\sin (T_{12}k_t) dk_t=
3T_{12}\tau_{12}\big[T_{12}^{2}-\tau_{12}^2\big]_{+}^{-5/2}~,\nonumber\\
\int_{0}^{\infty}k_tJ_0(\tau_{12}k_t)\cos (T_{12}k_t) dk_t=
\int_{0}^{\infty}k_t Y_0(\tau_{12}k_t)\sin (T_{12}k_t) dk_t=
-T_{12}\big[T_{12}^{2}-\tau_{12}^2\big]_{+}^{-3/2}~,
 \label{eq:E5.9}
\end{eqnarray}
where the distribution $x^{\lambda}_{+}$ is defined in a standard way with
the due number of subtracted terms of the Taylor expansion in the integral
$\int f(x) x^{\lambda}_{+}dx$ \cite{Gel'fand}.   Three different issues are
important here. First, each of the integrals (\ref{eq:E5.9}) is a well
defined distribution that includes all necessary regulators which provide
the convergence of subsequent integrations. The Bessel functions themselves
are {\em defined} as the Fourier transforms of the  (+)-distributions and we
just recover the original regular form by doing the inverse Fourier
transform (\ref{eq:E5.9}) (see Ref.\cite{Gel'fand})~. Second, as it will be
shown in the next section, the singular behavior  of the integrals
(\ref{eq:E5.9}) originates from the collinear domain. The (+)-prescription
that emerges here eliminates them term-by-term. Third, after the result of
the term-by-term integration is put back into  Eq.~(\ref{eq:E5.8}), the
singular collinear terms just cancel everywhere, including the null-plane
$\tau_{12}^2=0$. This type of cancelation of collinearly singular terms
takes place in all other pieces of the vacuum part of the quark self-energy.

All these observations lead us to the conclusion, that even in its vacuum
part, the self-energy does not suffer from collinear problems, which seems
to be a unique property of the expanding system. We do not continue to study
the vacuum part of the self-energy here, since we are currently interested
only in its material part which is discussed in the next section. [~The full
analysis of this part, including the issue of its renormalization, will be
published elsewhere.~]

\section{Radiation fields in the material part of the self-energy}
\label{sec:S6}

We found that the major contribution to the one-loop effective mass of a
``soft'' quark mode in the background of ``hard'' quarks and gluons comes
from the quark-quark forward scattering mediated by the magnetic component
of the longitudinal field. The purpose of this section is to demonstrate
that the interactions via transverse fields (including the forward
scattering of soft quark on hard gluons) is a secondary effect at least at
the very early stage of the nuclear collision. This conclusion may look
counter-intuitive, since, namely in the interactions of the transverse
fields, we expect to encounter the collinear enhancement of the radiation
amplitude. As it has been shown in Sec.\ref{sec:S5}, in the vacuum part of
the self-energy, the integrals of this type (taken in the limits from $0$ to
$\infty$) cancel each other leaving the vacuum sector free from collinear
divergences. The statistical weights ${\cal N}_g$ and ${\cal N}_f$, which
are different for the different terms, prevent such a cancelation in the
material part. Thus we have to analyze each term of the material part
separately.

As in Sec.\ref{sec:S5}, we consider an isolated piece which corresponds to
TE-gluons. The gluon correlators of this piece are the most singular and are
known not only in the integral representation, but in closed analytic form
also. The last circumstance is very helpful for the analysis of the multiple
integrals we meet below. (The terms identical to those computed below, also
appear in the part of self-energy due to the TM-gluons; the remaining terms
of TM-sector are less singular and, eventually, smaller than considered
here.) The corresponding fragment of the quark self-energy (\ref{eq:E3.13})
in the dispersion equation (\ref{eq:E3.12}) is
\begin{eqnarray} 
\bigg[\Sigma^{0}(\tau_1,\tau_2)\bigg]^{(TE)}_{mat}=
{i\alpha_s C_F\over 4\pi}~\theta(\tau_1-\tau_2)\int d^2 \vec{k_t}
\int_{-\infty}^{\infty}d\eta~ q_t \big[ g^{0}_{[0]} {\cal D}^{(TE)}_{[1]}
- g^{0}_{[1]} {\cal D}^{(TE)}_{[0]}\big] ,
\label{eq:E6.01}\end{eqnarray}
\begin{eqnarray}
\bigg[\Sigma^{T}(\tau_1,\tau_2)\bigg]^{(TE)}_{mat}=
{i\alpha_s C_F\over 4\pi}~
\theta(\tau_1-\tau_2)\int d^2\vec{k_t}\int_{-\infty}^{\infty}d\eta~
 \bigg[{(\vec{p_t} \vec{q_t})\over p_t^2} 
-2{(\vec{k_t} \vec{p_t})(\vec{k_t} \vec{q_t})\over k_t^2 p_t^2}\bigg]
\big[ g^{T}_{[0]}{\cal D}^{(TE)}_{[1]}-g^{T}_{[1]}{\cal D}^{(TE)}_{[0]}\big].
\label{eq:E6.02}\end{eqnarray}
The invariants of (anti-)commutators are the same as in the vacuum case,
\begin{eqnarray}
g^{0}_{[0]}(\tau_1,\tau_2;\eta;\vec{q_t})=
 i~{(\tau_1 -\tau_2)\cosh(\eta/2)\over 2\tau_{12}}
 \theta(\tau_{12}^2)~J_1(q_t\tau_{12})~,~~~~
 g^{T}_{[0]}(\tau_1,\tau_2;\eta;\vec{q_t})=
 -{\cosh(\eta/2)\over 2}
 \theta(\tau_{12}^2)~J_0(q_t\tau_{12})~,
\label{eq:E6.03}\end{eqnarray}
and the invariant ${\cal D}_{[0]}^{TE}(\tau_2,\tau_1;\eta;\vec{k_t})$ is
given by the first of Eqs.~(\ref{eq:E5.2a}). They all differ from zero only
for the timelike $\tau_{12}$. Hence, the material part of the invariants
$g_{[1]}$ and ${\cal D}^{(TE)}_{[1]}$ will be needed only in this domain. As
it has been discussed earlier, the distributions include only ``hard''
particles which are defined with respect  to the soft mode with the
transverse momentum $p_t$ by the inequalities,  $k_t>p_\ast$, and 
$q_t>p_\ast$, where $p_\ast\geq p_t$. Now, we are interested only in  the
material part with occupation numbers given by the equations,
\begin{eqnarray}
n_f(q_t,\theta)\approx {{\cal N}_f\over \pi R_\bot^2}
{\theta (q_t-p_\ast) \over q_t^2} ,~~~~
 n_g(k_t,\alpha)\approx {{\cal N}_g\over \pi R_\bot^2}
 {\theta (k_t-p_\ast) \over k_t^2} ~,
\label{eq:E6.04} 
\end{eqnarray}
and we must keep in mind the width $2Y$ of the rapidity plateau with the goal
to study if this is a significant parameter for the calculation of local
quantities. We may also question the validity of these formulae at
sufficiently large  $k_t$ and $q_t$, since without a cutoff, the integral
$\int dk_t/k_t$ diverges.

The material part of the densities will be employed in two different forms,
\begin{eqnarray}
g^{0}_{[1]}(\tau_1,\tau_2;\eta;\vec{q_t})= 
\int_{-\infty}^{\infty} {d\theta\over \pi}
n_f(\theta;q_t)\cosh\theta \sin q_t T_{12}(\theta)
=-~{(\tau_1 -\tau_2)\cosh(\eta/2)\over \tau_{12}}
n_f(q_t)~Y_1(q_t\tau_{12})~,
\label{eq:E6.05}
\end{eqnarray}
\begin{eqnarray}
g^{T}_{[1]}(\tau_1,\tau_2;\eta;\vec{q_t})= 
-~i\cosh{\eta\over 2}\int_{-\infty}^{\infty} {d\theta\over \pi}
n_f(\theta;q_t)~\cos q_t T_{12}(\theta)
=-~i\cosh(\eta/2)~n_f(q_t)~Y_0(q_t\tau_{12})
\label{eq:E6.06}
\end{eqnarray}
\begin{eqnarray}
{\cal D}_{[1]}^{(TE)}(\tau_2,\tau_1;\eta;\vec{k_t})= 
(\pi i)^{-1}\int_{-\infty}^{\infty} d\alpha 
n_g(\alpha;k_t) \cos k_t T_{12}(\alpha)
=i~Y_0(\tau_{12}k_t)~n_g(k_t)~,
\label{eq:E6.07}
\end{eqnarray}
where the second equation in (\ref{eq:E6.05})-(\ref{eq:E6.07}) is valid
only when $n_g$ and  $n_f$ are rapidity-independent, and we employ the
following notation,
\begin{eqnarray}
T_{12}(\alpha)=\tau_1\cosh(\alpha-\eta/2)-\tau_2\cosh(\alpha+\eta/2)=
\tau_{12}\cosh(\alpha -\psi)~,\nonumber\\
\tau_{12}^2=\tau_{1}^2+\tau_{2}^2 - 2\tau_1\tau_2\cosh\eta > 0,~~~~
\tanh\psi(\eta)=
{\tau_1+\tau_2\over\tau_1-\tau_2}\tanh{\eta\over 2},\nonumber\\ 
|\eta|<\eta_0 =\ln{\tau_1\over\tau_2}
\approx{\tau_1-\tau_2\over \sqrt{\tau_1\tau_2}}=\xi  , ~~~
\tanh\psi(\pm\eta_0)=\pm 1,~~~\psi(\pm\eta_0)=\pm \infty~, 
\label{eq:E6.08} 
\end{eqnarray} 

where $\xi=(\tau_1-\tau_2)/\sqrt{\tau_1\tau_2}\approx\eta_0$ is the main
parameter  of our calculations. This parameter is supposed to be small in
order that the notion of the current transverse mass $\mu(p_t,\tau_1)$ has
the expected meaning of a slowly varying parameter. The geometric mean time
$\tau_m= \sqrt{\tau_1\tau_2}$ has a simple interpretation. The two
characteristics, one connecting the points $(\tau_2,-\eta_0)$ and 
$(\tau_1,\eta_0)$, and the second one connecting the points
$(\tau_2,\eta_0)$ and  $(\tau_1,-\eta_0)$, intersect at the point
$(\tau_m,0)$. The proper time $\tau_m$ is always inside the domain of the
``causal interaction.''

Let us start the analysis of the radiation-dominated terms with the invariant
$[\Sigma^{0}]^{(TE)}_{mat}$.  (The invariant $[\Sigma^{T}]^{(TE)}_{mat}$
appears to have an extra small factor $\xi$.) According to
Eqs.~(\ref{eq:E6.04}), (\ref{eq:E6.05}), and (\ref{eq:E6.07}) it can be
written as a multiple integral, 
\begin{eqnarray}
\bigg[\Sigma^{0}(\tau_1,\tau_2)\bigg]^{(TE)}_{mat}=
{i\alpha_sC_F\over 8\pi^2}
\int_{-\infty}^{+\infty} d\eta  \theta(\tau_{12}^2) 
\bigg\{{{\cal N}_f\over \pi R_\bot^2}\int_{p_\ast}^{\infty} d q_t
\int_{0}^{2\pi}d\varphi ~J_0(\tau_{12}k_t)
\int_{-\infty}^{\infty}d \theta~\cosh \theta
\sin T_{12}(\theta)q_t \nonumber\\
+{{\cal N}_g\over \pi R_\bot^2}
{(\tau_1-\tau_2)\cosh\eta/2\over\tau_{12}}
\int_{p_\ast}^{\infty} {d k_t\over k_t}\int_{0}^{2\pi}d\varphi
~q_t J_1(\tau_{12}q_t)\int_{-\infty}^{\infty}d\alpha~
\cos T_{12}(\alpha)k_t ~\bigg\}\theta(\tau_1-\tau_2)~,
\label{eq:E6.09} 
\end{eqnarray}
where we choose the integral form of the densities $g_{[1]}$ and  ${\cal
D}^{(TE)}_{[1]}$ in order to find the domain in the multidimensional space
where the dominant contribution comes from. Since the two terms in
(\ref{eq:E6.09}) are not expected to interfere (or UV- diverge), we are free
to change variables in these terms independently. We leave $d^2\vec{k_t}$ in
the second term, and change for  $d^2\vec{q_t}$ in the first one. The next
step is to integrate over the azimuthal angle between $\vec{q_t}$ and
$\vec{p_t}$ in the first term of Eq.~(\ref{eq:E5.3}), and over the angle
between $\vec{k_t}$ and $\vec{p_t}$ in the second term. This integration
deals only with the retarded propagators inside the self-energy loop and
selects the lowest angular harmonics,
\begin{eqnarray}
\int_{0}^{2\pi}q_t J_1(\tau_{12}q_t)d\varphi=
2\pi~[ k_t J_0(\tau_{12}p_t)J_1(\tau_{12}k_t)+
p_t J_1(\tau_{12}p_t)J_0(\tau_{12}k_t)]~,~~~k_t>p_t~,
\label{eq:E6.10a}
\end{eqnarray}
\begin{eqnarray}
\int_{0}^{2\pi}~ J_0(\tau_{12}k_t)d\varphi=
2\pi~ J_0(\tau_{12}p_t)J_0(\tau_{12}q_t)~,~~~q_t>p_t~.
\label{eq:E6.10b}
\end{eqnarray}
Only the first of the two terms in Eq.~(\ref{eq:E6.10a}), corresponding to 
the collinear geometry in the transverse plane survives in the limit of 
$k_t\gg p_t$ and has to be retained by our major assumption. The second term
describes the deviation from collinearity and is small. However, it is
instructive to keep it for a while.  After these angular integrations,
Eq.~(\ref{eq:E6.09}) becomes
\begin{eqnarray}
\bigg[\Sigma^{0}(\tau_1,\tau_2)\bigg]^{(TE)}_{mat}=
{i\alpha_sC_F\over 4\pi}\theta(\tau_1-\tau_2)
\int_{-\infty}^{+\infty} d\eta  \theta(\tau_{12}^2) 
\bigg\{{{\cal N}_f\over \pi R_\bot^2}\int_{p_\ast}^{\infty} d q_t
J_0(\tau_{12}p_t)J_0(\tau_{12}q_t)
\int_{-\infty}^{\infty}d \theta~\cosh \theta
\sin T_{12}(\theta)q_t \nonumber\\
+{{\cal N}_g\over \pi R_\bot^2}
{(\tau_1-\tau_2)\cosh\eta/2\over\tau_{12}}
\int_{p_\ast}^{\infty} {d k_t\over k_t}
[ k_t J_0(\tau_{12}p_t)J_1(\tau_{12}k_t)
+p_t J_1(\tau_{12}p_t)J_0(\tau_{12}k_t)]
\int_{-\infty}^{\infty}d\alpha 
\cos T_{12}(\alpha)k_t ~\bigg\},
\label{eq:E6.11} 
\end{eqnarray}
where the actual limits of integration over $\eta$, $\theta$, and $\alpha$ 
have yet to be put in agreement with the model we employ. Now, we have
approached the most subtle point of our analysis. This expression includes
triple integrations, any of which (if performed formally) yields singular
functions.  For the sake of definiteness, let us start with the second
term in Eq.(\ref{eq:E6.11}) (which corresponds to the forward scattering of
soft quark on a hard gluon from the distribution $n_g(\alpha,k_t)$), 
rewriting it in its most expanded form,
\begin{eqnarray}
{\cal J}_2={{\cal N}_g\over \pi R_\bot^2}\int d\eta \theta(\tau_{12}^2) 
{(\tau_1-\tau_2)\cosh\eta/2\over\tau_{12}}J_0[\tau_{12}(\eta)p_t]
\int_{-\infty}^{\infty}d\alpha
\int_{p_\ast}^{\infty}J_1[\tau_{12}(\eta)k_t]
\cos [k_t\tau_{12}(\eta)\cosh(\alpha-\psi(\eta))]~d k_t~.
\label{eq:E6.12} 
\end{eqnarray}

\begin{figure}[htb]
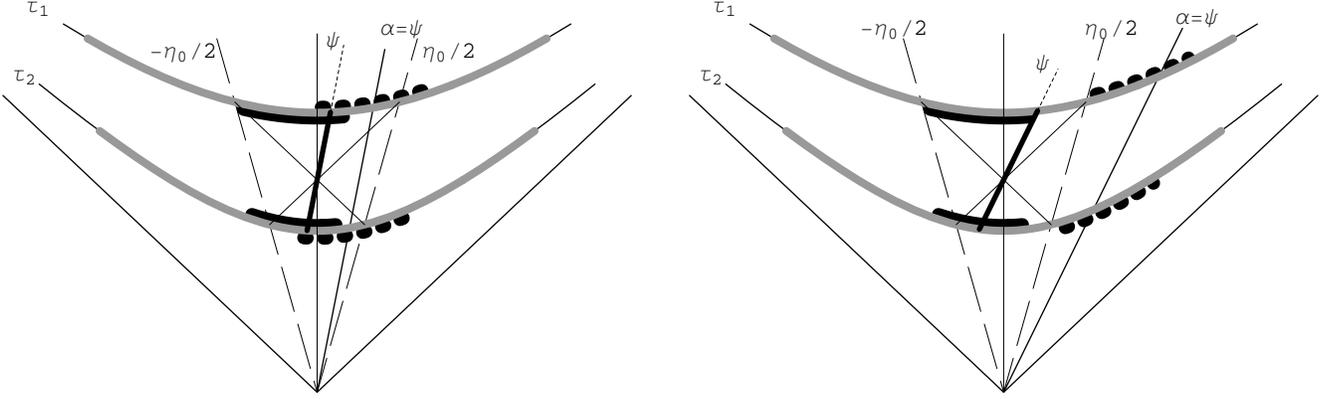

\begin{center}
\mbox{ 
\psfig{file=./col1.ps,height=2.2in,bb=105 535 385 715}
\hspace{.2cm}
\psfig{file=./col2.ps,height=2.2in,bb=105 535 385 715}
}
\end{center}
\caption{Geometry of fields in the forward scattering amplitude $qg\to qg$.}
\label{fig:fig2}
\end{figure}

The first observation is that at large $k_t$ (which is the condition that
the distribution $n_g(\alpha,k_t)$ can be measured within a short time) the
main contribution to the $\alpha$-integration comes from the domain
$\alpha\approx\psi(\eta)$ where the phase of the $\cos T_{12}(\alpha)k_t$ is
stationary.  This is the domain of collinear interaction when the hard gluon
from the distribution $n_g(\alpha,k_t)$ has almost the same rapidity as the
virtual quark with transverse momentum $~\vec{q_t}=\vec{k_t}+\vec{p_t}~$ in
the self-energy loop. Obviously, this quark is also hard. Furthermore, its
propagator, $G_{[ret]}(\tau_1,\tau_2;q_t)= \theta
(\tau_1-\tau_2)G_{[0]}(\tau_1,\tau_2;q_t)$, is devised only from the free
on-mass-shell partial waves which themselves are well localized in the
rapidity direction. Hence, we deal with the intuitively very clear case of
collinear absorption and emission of the gauge field quantum by a charged
particle. All participants of the process are moving with the same
velocity.  According to  the property of localization of states in wedge
dynamics studied in paper [II], such a fine tuning of  $\alpha$ to $\psi$ 
is indeed possible.  This  is illustrated by the left-hand figure on 
Fig.\ref{fig:fig2}. where the grey segments of the hyperbolas $\tau=\tau_2$
and $\tau=\tau_1$ correspond to the rapidity intervals occupied by the soft
quark mode, $\tau p_t<1$, at the beginning and at the end of the scattering
process, respectively. The bold black and the dashed segments  show the 
rapidity intervals where  the hard virtual quark and the hard gluon are
localized at the same times. All three fields effectively overlap around
rapidity $\eta_2=-\eta/2$ at $\tau=\tau_2$ and around $\eta_1=+\eta/2$ at
$\tau=\tau_1$. The rapidity direction between these points is exactly
$\psi(\eta)$.  The rapidity  $\alpha$ of the external gluon is sufficiently
small and is close to the rapidity $\psi(\eta)$.  

The maximal rapidity width of the interaction domain is defined by the
causality condition $\tau_{12}^2 >0$, which immediately establishes the
upper boundary  $|\eta|<\eta_0$.   Since the collinear interaction
corresponds to the  condition $\alpha\approx\psi(\eta)$, the rapidity of the
hard gluon must be within  this geometrically defined interval as well. The
opposite case is depicted in the right-hand figure of Fig.\ref{fig:fig2}.
The rapidity $\psi(\eta)$ is so large, that the external gluon is not
localized within the causal boundaries $\pm\eta_0/2$ of the interaction
domain. In order to avoid this, we have to impose an  even stronger
requirement that $|\psi(\eta)|<\eta_0$.  According to  (\ref{eq:E6.08}), we
have $|\psi(\eta)|>\eta$. Hence, we must further take  $|\eta|<\eta_\ast$,
where the boundary $\eta_\ast$ is defined by the equation
$\psi(\eta_\ast)=\eta_0$, 
\begin{eqnarray}
{\tau_1+\tau_2\over\tau_1-\tau_2}\tanh{\eta_\ast\over 2}=\tanh\eta_0
\equiv {\tau_1^2-\tau_2^2\over\tau_1^2+\tau_2^2}~,
\label{eq:E6.13} 
\end{eqnarray}
which has a solution,
\begin{eqnarray}
\tanh{\eta_\ast\over 2}
= {(\tau_1-\tau_2)^2\over\tau_1^2+\tau_2^2}~,~~~~~~
\eta_\ast\approx{(\tau_1-\tau_2)^2\over \tau_1\tau_2}~=\xi^2~.
\label{eq:E6.14} 
\end{eqnarray}
We remind the reader that $\xi\ll 1$; only this condition allows one to 
introduce the the time-dependent transverse mass $\mu(p_t,\tau)$.
In order to simplify further analysis, it is convenient to present the
internal integral over $k_t$  as the difference,
\begin{eqnarray}
 -~{\tau_{12}\over (T^2_{12}(\alpha)-\tau_{12}^2)_+^{1/2}
[T_{12}(\alpha)+(T^2_{12}(\alpha)-\tau_{12}^2)^{1/2}]}~
-\int_{0}^{p_\ast} J_1(\tau_{12}k_t)
\cos [T_{12}(\alpha)k_t]~d k_t ~,
\label{eq:E6.15} 
\end{eqnarray}
where the first singular term is the integral over $k_t$, computed from $0$ 
to $\infty$, and thus, it completely accounts for the domain
$k_t\to\infty$. It includes the function 
\begin{eqnarray}
f(\eta,\alpha) =[T^2_{12}(\alpha)-\tau_{12}^2]_+^{-1/2}=
[\tau_{12}^2(\eta)\sinh^2(\alpha-\psi(\eta))]_+^{-1/2}~, \nonumber
\end{eqnarray}
which is singular at $\alpha = \psi(\eta)$, thus fully accounting for the
expected collinear enhancement. This function, however,
is  a canonical distribution with respect to both its arguments $\alpha$
and $\eta$, and it carries the standard regulators for the subsequent
integrations. We shall consider the singular and the regular terms 
separately. Using the above found limits, we may write the singular term as
\begin{eqnarray}
I_{2}^{sing}=\int_{-\eta_\ast}^{\eta_\ast} d\eta 
{(\tau_1-\tau_2)\cosh\eta/2\over\tau_{12}(\eta)}
{1\over [\tau_{12}(\eta)]_+ }~
\int_{-\eta_0}^{\eta_0}d\alpha~
{e^{-|\alpha-\psi |}\over [\sinh^2|\alpha -\psi |]_{+}^{1/2}}~,
\label{eq:E6.16} 
\end{eqnarray}
where, since $\tau_{12}p_t\ll 1$, we put $J_0[\tau_{12}p_t]\approx 1$. After
an obvious change of variable, the internal integral of Eq.~(\ref{eq:E6.16})
can be split  into two,
\begin{eqnarray}
\int_{-\eta_0}^{\eta_0} d\alpha~
{e^{-|\alpha-\psi |}\over [\sinh^2|\alpha -\psi |]_{+}^{1/2}}=
\bigg[ \int_{0}^{\eta_0+\psi}+\int_{0}^{\eta_0-\psi}\bigg] 
{\alpha~e^{-\alpha}\over \sinh\alpha }~{d\alpha\over \alpha_{+}}~.\nonumber
\end{eqnarray}
Since by the definition of the (+)-distribution, 
\begin{eqnarray}
\int_{0}^{\beta} {\alpha~e^{-\alpha}\over \sinh\alpha }
~{d\alpha\over \alpha_{+}}= 
\int_{0}^{\beta}{e^{-\alpha }d\alpha\over\sinh\alpha}
-\int_{\epsilon}^{1}{d\alpha\over\alpha}=
\ln{1-e^{-2\beta}\over 2}~,
\label{eq:E6.17} 
\end{eqnarray}
we obtain the singular part in the form,
\begin{eqnarray}
I_2^{sing}=2~\int_{0}^{\eta_\ast}
{(\tau_1-\tau_2)\cosh\eta/2\over\tau_{12}(\eta)}
{ d\eta \over [\tau_{12}(\eta)]_+ }~
\ln\big\{{1\over 4}[1+e^{-4\eta_0}-2e^{-2\eta_0}\cosh 2\psi(\eta)]\big\}~.
\label{eq:E6.18} 
\end{eqnarray}
Next, it is convenient to trade $\eta$ for a new variable $y$, 
$\tau_{12}(\eta)=(\tau_1 -\tau_2)y$. The helpful relations for this change
of variables are
\begin{eqnarray}
{\cosh(\eta/2)d\eta \over\tau_{12}(\eta)}={-1\over\sqrt{\tau_1\tau_2}}~
{d y\over \sqrt{1-y^2}}~,~~~~
\tau_{12}^2(\eta_\ast)=
{(\tau_1 -\tau_2)^2\over 1+(\tau_1 -\tau_2)^2/\tau_1\tau_2}~,\nonumber\\
y_\ast\approx 1-{(\tau_1 -\tau_2)^2\over 2\tau_1\tau_2}~,~~~~
\cosh 2\psi={(\tau_1 +\tau_2)^2\over 2\tau_1\tau_2}{1\over y^2}-
{\tau_1^2 +\tau_2^2\over 2\tau_1\tau_2}~.\nonumber
\end{eqnarray}
Taking into account that $\cosh 2\psi\to 1$, when $y\to 1$, we obtain,
\begin{eqnarray}
I_2^{sing}={4\over\sqrt{\tau_1\tau_2}}~\int_{y_\ast}^{1}
{d y\over y\sqrt{1-y^2}}
\ln{1-e^{-2\eta_0}\over 2} \approx {4\over\sqrt{\tau_1\tau_2}}
~{\tau_1-\tau_2\over\sqrt{\tau_1\tau_2}}
~\ln{\tau_1-\tau_2\over\sqrt{\tau_1\tau_2}}=
{4\over\sqrt{\tau_1\tau_2}}~\xi\ln\xi~.
\label{eq:E6.19} 
\end{eqnarray}

This formula has two distinctive elements. The first element is the large
$\ln\xi$, which is due to the  collinear geometry of the interaction. This
would lead to a divergence if the interaction domain were unlimited. The
second element is the small factor $\xi$ which is due to the small volume
occupied by the  interaction and it completely suppresses the potential
divergence.  One may notice that when the mean time  $\sqrt{\tau_1\tau_2}$
increases, then $\xi\to 0$, and the corresponding part of the self-energy
also tends to zero. This is easy to understand, since with the mean time
growing, the system becomes more and more diluted locally.

The regular part of Eq.~(\ref{eq:E6.12}) is given by the integral,
\begin{eqnarray}
I_2^{reg}=\int_{-\eta_\ast}^{\eta_\ast} d\eta 
{(\tau_1-\tau_2)\cosh\eta/2\over\tau_{12}(\eta)}
\int_{-\eta_0}^{\eta_0}d\alpha~
\int_{0}^{p_\ast} J_1(\tau_{12}k_t)
\cos [T_{12}(\alpha)k_t]~d k_t ~,
\label{eq:E6.20} 
\end{eqnarray}
where, when $~-\eta_0<\eta<\eta_0$, then $T_{12}(\alpha)$ varies between
its minimal and maximal values,
\begin{eqnarray}
{(\tau_1 -\tau_2)^2\over\sqrt{\tau_1\tau_2} }e^{-|\alpha|} <
T_{12}(\alpha)<{(\tau_1 -\tau_2)^2\over
\sqrt{\tau_1\tau_2} }e^{+|\alpha|}~.      \nonumber
\end{eqnarray}
 Therefore, when $k_t<p_t$,
we have $T_{12}k_t\sim (\tau_1-\tau_2)k_t\ll 1$, and both functions under
the integral over $k_t$ can be expanded in Taylor series. All integrations
become trivial and yield \footnote{ Even if we impose no limitations on
$\alpha$ the estimate is still as small as 
$${\tau_1\tau_2 p_\ast^2\over\sqrt{\tau_1\tau_2}}~\xi^2\ln\xi~.$$}
\begin{eqnarray}
I_2^{reg}= {\tau_1\tau_2 p_\ast^2\over\sqrt{\tau_1\tau_2}}~
\bigg[{\tau_1-\tau_2\over\sqrt{\tau_1\tau_2}}\bigg]^4
={\tau_1\tau_2 p_\ast^2\over\sqrt{\tau_1\tau_2}}~\xi^4~.
\label{eq:E6.21} 
\end{eqnarray}
We have chosen this form of the answer, because our major assumption is 
valid only as long as $\tau p_t \leq 1$ and because the dispersion equation 
(\ref{eq:E3.12}) has a kinematic factor $\sqrt{\tau_1\tau_2}$ in it.

The other two terms in Eq.~(\ref{eq:E6.11}) can be studied along the same
guidelines. The third term is suppressed with respect to ${\cal J}_2$ by the
factor $p_t/k_t$, which is small by our major model agreement and it could
have been discarded on this ground only. To be on safe side, let us rewrite
it as
\begin{eqnarray}
{\cal J}_3={{\cal N}_g\over \pi R_\bot^2}\int_{-\eta_\ast}^{\eta_\ast} 
d\eta ~{(\tau_1-\tau_2)\cosh\eta/2\over\tau_{12}}
~p_t~J_1[\tau_{12}(\eta)p_t]\nonumber\\
\times\int_{-\eta_0}^{\eta_0} d\alpha ~ 
\bigg\{ -\gamma_E -\ln\big[{\tau_{12}p_t\over 2}
e^{|\alpha-\psi(\eta)|}\big]+
\int_{0}^{p_\ast}{1- J_0[\tau_{12}(\eta)k_t]\cos (T_{12}(\alpha)k_t)
\over k_t}~d k_t~\bigg\}~,
\label{eq:E6.22} 
\end{eqnarray}
where the first term is the integral over $k_t$ from $0$ to $\infty$. As i
could be expected, the integrand is regular. Since there is an accounted for
difference between $k_t$ and $q_t$, the exactly collinear regime becomes
impossible and we do not have the large collinear logarithm in ${\cal J}_3$.
Overall, this term is also suppressed at least by a  factor $\xi$ stemming 
from $J_1[\tau_{12}(\eta)p_t]$ in the integrand.

The first term in Eq.(\ref{eq:E6.11}) corresponds to the forward quark-quark 
scattering with high momentum transfer,
\begin{eqnarray} {\cal J}_1={{\cal N}_f\over \pi R_\bot^2}
\int_{-\eta_\ast}^{\eta_\ast}  d\eta 
~J_0[\tau_{12}(\eta)p_t] \hspace{10cm}\nonumber\\ 
\times \int_{-\eta_0}^{\eta_0} d\theta ~  
\bigg\{ {\cosh\theta\over [\sinh^2(\theta-\psi(\eta))]_{+}^{1/2}} 
{1\over[\tau_{12}(\eta)]_+}
-\cosh\theta~\int_{0}^{p_\ast} J_0[\tau_{12}(\eta)q_t] 
\sin (T_{12}(\theta)q_t)~d q_t~\bigg\}~. 
\label{eq:E6.23}  \end{eqnarray} 
Here, we again recognize the collinear singularity which is, as previously,
regulated by the (+)-prescription. All further  calculations for 
${\cal J}_1$ are similar to the case of ${\cal J}_2$ and the answer reads, 
\begin{eqnarray} 
I_1^{sing} \leq {-8\xi + 4 \xi^2 \ln\xi\over \sqrt{\tau_1\tau_2}}~,~~~~ 
I_1^{reg}\approx 2{p_\ast^2 \tau_1\tau_2 \over \sqrt{\tau_1\tau_2}}\xi^5~. 
\label{eq:E6.24} 
\end{eqnarray} 
These results will serve for us as a reference point for the estimates of
the mathematically more complicated part connected with the radiation field
of the transverse magnetic TM-modes. Before we address this issue, it is
expedient to look at the obtained results more attentively and trace the
correspondence between the calculations and physical picture in more
details: 

{\em ~~1.} It has been observed in Sec.~\ref{sec:S5} (for the vacuum part of
the quark self-energy) that in the framework of wedge dynamics, the
collinear problems do not jeopardize the field theory. In ``material part''
of the self-energy, the collinear interactions were proved to be the most
intensive and to lead to a  visible enhancement of the interaction between
the quark and {\em radiation} field. However, this enhancement never turns
into a disaster of collinear divergence. One of the trivial reasons is that
the space-time domain of the interaction is now limited, and large
logarithms are multiplied by small phase volumes.  

{\em ~~2.} A deeper insight into the  wedge dynamics, shows that even
intermediate  collinear singularities observed in the terms ${\cal J}_1$
and  ${\cal J}_2$ are spurious. In order to reveal this fact, one can notice
that the  singularity at $\alpha = \psi(\eta)$ is present only in ${\cal
J}_1$ and ${\cal J}_2$.  It is absent in ${\cal J}_3$,  because of the extra
negative power of $k_t$ brought by the subleading term of the angular
integral $d\varphi$. This extra $k_{t}^{-1}$ effectively suppresses the
distribution $n_g(\alpha,k_t)$ at large $k_t$.  Next, one may ask, what
minimal change of $n_g(\alpha,k_t)$ at large $k_t$ is necessary in order
that the intermediate collinear singularity does not appear at all. This can
be learned by  changing the order of integration in Eq.~(\ref{eq:E6.12}).
One can start from the integral $d\eta$ with an assumption that the
integrand only slowly varies within some interval of $\alpha$ around $\alpha
=0$. Then, it is easy to see that the singular term $I_2^{sing}$ of
Eq.~(\ref{eq:E6.19}) {\em totally} originates from the domain of
$k_t\to\infty$. It comes from a logarithmic integral between two infinite
limits.  This residual piece emerges only because we extend the distribution
$n_g\sim k_t^{-2}$ (obtained from a dimensional estimate) to an arbitrarily
large $k_t$. As we have already mentioned, this dependence is unphysical,
e.g., because the distribution $d^2\vec{k_t}/k_t^2$ is not normalizeable. It
has to be modified above some value of $k_t$ and, therefore, the singular
term must vanish completely.  Thus, in wedge dynamics, the phenomenon of
collinear enhancement is intrinsically connected with the basic property of
localization inherent in the one-particle states. Only the states with
infinitely large $k_t$ can have a precisely given rapidity and be
responsible for the singularities like we encounter in Eqs.~(\ref{eq:E6.15})
and  (\ref{eq:E6.23}).

{\em ~~3.}  Our  way to pick out the leading contributions (in the mixed
representation of wedge dynamics) from the space-time domains, where the
phases of the interacting fields are stationary, is a generalization of the
known method of isolating the leading terms using the pinch-poles in the
plane of complex energy. The similarity of two methods can be easily
understood since, e.g., the quark density correlator in the self-energy can
be presented as a sum of two propagators,
$G_{[1]}(q)=G_{[00]}(q)+G_{[11]}(q)$. In the plane of the complex energy
$q^0=k^0+p^0$, the (Feynman-type) propagator $G_{[00]}(q)$ has poles in the
second and fourth quadrants, while the (anti-Feynman-type) propagator
$G_{[11]}(q)$ has poles in the first and third quadrants.  The radiation
part of the retarded gauge field propagator $D_{[ret]}(k)$ has poles in the 
third and fourth quadrants. Therefore, in both terms of  $G_{[1]}(q)
D_{[ret]}(k) =G_{[00]}(q)D_{[ret]}(k) +G_{[11]}(q)D_{[ret]}(k)$, the
integration path along the real axis of the complex $k^0$ plane is pinched
between two poles (one from $D_{[ret]}$, and the second from $G_{[00]}$ or 
$G_{[11]}$ ) giving the leading contribution when $p^0$ is small, and the
three-momenta $\bbox{k}$ and $\bbox{q=k+p}$ are are very close to each
other. Similar arguments are valid for the second part, $G_{[ret]}D_{[1]}$,
of the quark self-energy. The term $G_{[1]}(q)D_{[long]}(k)$ is exceptional,
because the propagator $D_{[long]}(k)$ of the longitudinal field has no
poles corresponding to the propagation.

The wedge dynamics does not allow for a standard momentum representation,
since its geometric background is not homogeneous in the $t$- and
$z$-directions; accordingly, we do not have familiar pinch-poles in our
calculations. Nevertheless, the patches of phase space where the phases of
certain field fragments are stationary and effectively overlap, do now the
same job as the pinch-poles, and yield the same answers when the homogeneity
required for the momentum representation is restored. The way wedge dynamics
tackles the problem is genuinely more general, because it addresses the
space-time picture of the interacting fields.\footnote{It is well known,
that the threshold behavior of the imaginary part of the photon self-energy
can be derived from  the pinch geometry of the poles of the electron
propagators \cite{Berest}. Since at the threshold, the $e^+ e^-$ pair is
created with zero relative velocity, the pinch indeed corresponds to the
overlap of the stationary phases of the $e^+$ and $ e^-$ wave functions in
the maximal possible volume.} The momentum space is now split into the 
subspaces of rapidity and transverse momentum; the correlation between the 
particle's rapidity and its location is increasing with the increase of its
transverse momentum . The role of pinch-poles is taken over by the
geometrical overlap of the field patterns with the same rapidity. This
observation can serve as a footing for the future development of  an
effective technique for perturbative calculations in wedge dynamics. The
arguments of the localization are not applicable to the  longitudinal part
of the gluon field. In the term $G_{[1]}D_{[long]}$, no patch in space-time
is dynamically selected, since $D_{[long]}$ is not assembled from the
propagating waves that could match the virtual quark in the loop by their
phase. This is in line with the absence of pinch-poles due to
$D_{[long]}(k)$ in the momentum picture.

The arguments presented above allow one to estimate the contribution of the
radiation fields of the TM-mode in a very economical way. Let us consider
the group ${\cal D}_{[0]}^{(2)}g^0_{[1]}$, which is very similar to the term
${\cal J}_{1}$ studied above, as an example. Now, the invariant  ${\cal
D}_{[0]}^{(2)}$, as can be inferred from Eq.~(\ref{eq:A1.4}), is known only
in the integral representation, and not in an analytic form. Let us,
therefore, employ the analytic form of $g^0_{[1]}$, given by
Eq.~(\ref{eq:A1.20}). It is easy to see, that the integration over $\alpha$
in ${\cal D}_{[0]}^{(2)}$ leads to the same causal step-function 
$\theta(\tau_{12}^2)$ and, as previously shown,  we have $|\eta|<\eta_0$.
The expanded form of this term is
\begin{eqnarray}
{\cal J}_4={{\cal N}_f\over \pi R_\bot^2}\int_{-\eta_0}^{\eta_0}
 d\eta \theta(\tau_{12}^2) 
{(\tau_1-\tau_2)\cosh\eta/2\over\tau_{12}}\int d\alpha 
\tanh\big(\alpha-{\eta\over 2}\big)  \tanh\big(\alpha+{\eta\over 2}\big)
\int_{p_\ast}^{\infty} Y_1[\tau_{12}(\eta)q_t]
\sin [ T_{12}(\alpha)q_t]dq_t ,
\label{eq:E6.25} 
\end{eqnarray}
where we have integrated out the azimuthal angle $\varphi$ assuming that 
$k_t,q_t\gg p_t$ (the first correction is smaller by the factor 
$(p_t/k_t)^2\ll 1$). The internal integral over $q_t$ can be transformed into
\begin{eqnarray}
 -~{\cosh (\alpha-\psi(\eta))]\over 
 [\tau_{12}^2(\eta)\sinh^2(\alpha-\psi(\eta))]_+^{1/2}}~
-~\int_{0}^{p_\ast} Y_1(\tau_{12}q_t) \sin [T_{12}(\alpha)k_t]~d k_t ~,
\label{eq:E6.26} 
\end{eqnarray}
which brings us very close to Eq.~(\ref{eq:E6.23}) for ${\cal J}_1$. Once
again, we encounter a collinear singularity at $\alpha=\psi(\eta)$, and
exactly the same arguments force us to set the same limits in the integrals,
as in  Eq.~(\ref{eq:E6.23}). We do not have to continue the calculations to
understand the smallness of ${\cal J}_4$, mention only, that due to the
narrow limits of two rapidity integrations in Eq.~(\ref{eq:E6.25}), the
product of the two hyperbolic tangents in the integrand will add extra
$\xi^2$ to the order of smallness of ${\cal J}_4$. By the same argument as
used previously, we can easily  learn that the singular term in
Eq.~(\ref{eq:E6.26}) is spurious.

In this group, associated with the $TM$-mode of the radiation field, the
leading (still parametrically small, and equal to ${\cal J}_1$) 
contribution comes from the  $D^{\eta\eta}$ component of the gluon
correlators. We may summarize by the observation that only the overlap of the
domains of stationary phase in two correlators matters. It can be visualized
via the partial-wave expansion of any of the correlators in the self-energy
loop.

\section{ Nonlocal part of the longitudinal propagator 
     in the material part of the self-energy.}
\label{sec:SA2}

The longitudinal part of the gluon propagator contributes, to the invariant
$\Sigma^{0}$, the term
\begin{eqnarray} 
\bigg[\Sigma^{0}(\tau_1,\tau_2)\bigg]^{[long]}_{mat}=
{i\alpha_s C_F\over 4\pi}~\theta(\tau_1-\tau_2)\int d^2 \vec{k_t}
\int_{-\infty}^{\infty}d\eta~ q_t ~g^{0}_{[1]}
\bigg[{\cal D}^{(2)}_{[long]} 
+{1\over \tau_1\tau_2}{\cal D}^{(\eta\eta)}_{[long]}\bigg]~.
\label{eq:A2.01}\end{eqnarray}
Using Eqs.~(\ref{eq:A1.14a}) and (\ref{eq:A1.15a}) for the gluon invariants,
and the first of Eqs.~(\ref{eq:A1.20}) for the quark invariant $g^{0}_{[1]}$,
we arrive at the following expression which accounts for the non-local part
of the longitudinal propagator (the contact part was studied in 
Sec.~\ref{sec:S3a}).
\begin{eqnarray} 
\bigg[\Sigma^{0}(\tau_1,\tau_2)\bigg]^{[long]}_{mat}=
{i\alpha_s C_F{\cal N}_f\over 2\pi^2R_\bot^2}~\theta(\tau_1-\tau_2)~
\int_{p_\ast}^{\infty} d q_t
\int_{-\infty}^{\infty}d\eta~ 
{(\tau_1-\tau_2)\cosh\eta/2\over |\tau_{12}^2|^{1/2}} \nonumber\\
\times {k_t\cosh \eta\over 2} \int_{\tau_2}^{\tau_1} e^{-t q_t\sinh |\eta|} 
\bigg(1-{t^2\over\tau_1\tau_2}\bigg) ~ dt 
\bigg[\theta(\tau_{12}^2)~Y_1(\tau_{12}q_t)
+{2\over\pi}\theta(-\tau_{12}^2)~
K_1(\tilde{\tau}_{12}q_t)\bigg] ,
\label{eq:A2.02}\end{eqnarray}

where we have integrated out the dependence on the azimuthal angle 
$\varphi$ in the approximation of $q_t\gg p_t$. The key observation that
allows one to judge about the smallness of this term is that the limits of
integration over variable $t$ are very close, and the factor 
$[1-t^2/\tau_1\tau_2]$ is very small. This factor reflects a known
competition between the electric and magnetic interaction of  moving charges
which reduces the net yield almost to zero.  Let us replace $t$ by the
dimensionless  $u=t/\sqrt{\tau_1\tau_2}$. Then this part of the integral
becomes 
\begin{eqnarray} 
{1\over\sqrt{\tau_1\tau_2}}
\int_{\tau_2}^{\tau_1}
\bigg[1-{t^2\over\tau_1\tau_2}\bigg]\cdot\cdot\cdot dt
= \int_{\sqrt{\tau_2/\tau_1}}^{\sqrt{\tau_1/\tau_2}}
(1-u^2)\cdot\cdot\cdot du
=\int_{\sqrt{1-\xi^2/4}-\xi/2}^{\sqrt{1-\xi^2/4}+\xi/2}
(1-u^2)\cdot\cdot\cdot du \approx -{\xi^3\over 3}f(\xi)~,
\label{eq:A2.03}\end{eqnarray}
where the limiting behavior $f(\xi)\sim const/\xi$  when $\xi\to 0$ can be
conjectured from the behavior of the functions $Y_1(x)$ and $K_1(x)$ at
small $x$. However, the Laplace transforms of these functions in
Eq.~(\ref{eq:A2.02}) are singular functions of $\eta$ and we have to be
careful in estimating these terms. In fact, the $\xi^2$ order of smallness
is not altered by the remaining integrations. First, it is useful to notice
that the last factor in square brackets in Eq.~(\ref{eq:A2.02}) is nothing
but the invariant $g^{0}_{[1]}$ which, according to its integral
representation (\ref{eq:A1.20}),  equals zero at $\xi=0$.  Next, it is
profitable to change the variables of integration in the following way. We
trade $\eta$ for $y$ according to $\tau_{12}(\eta)=(\tau_1 -\tau_2)y$ in the
domain $\tau_{12}^2>0$. In the complimentary domain
$\tau_{12}^2=-\tilde{\tau}_{12}^2<0$, we change $\eta$ for $y$ using
$\tilde{\tau}_{12}(\eta)=(\tau_1 -\tau_2)y$. We also replace $\xi q_t$ by a
new variable $q$.
\begin{eqnarray}
\bigg[\Sigma^{0}(\tau_1,\tau_2)\bigg]^{[long]}_{mat}=
{i\alpha_s C_F{\cal N}_f\over 2\pi^2R_\bot^2}~\theta(\tau_1-\tau_2)
~{1\over\xi}~\int_{\sqrt{1-\xi^2/4}-\xi/2}^{\sqrt{1-\xi^2/4}+\xi/2} (1-u^2)
\int_{\xi p_\ast}^{\infty} q d q \nonumber \\
\times\bigg\{\int_0^1 {dy\over\sqrt{1-y^2}}
\bigg[1+{\xi^2\over 2}(1-y^2)\bigg]
~{\pi\over 2}~Y_1(\tau_m qy) 
~e^{-\tau_m q u\sqrt{1-y^2}~\sqrt{1*\xi^2(1-y^2)/2}}\nonumber \\
+\bigg(\int_0^1+\int_1^\infty\bigg)
 {dy\over\sqrt{1+y^2}}\bigg[1+{\xi^2\over 2}(1+y^2)\bigg]
K_1(\tau_m qy) 
~e^{-\tau_m q u\sqrt{1+y^2}~\sqrt{1*\xi^2(1+y^2)/2}}\bigg\}~.
\label{eq:A2.04}\end{eqnarray}
where, we remind the reader that $\tau_m=\sqrt{\tau_1\tau_2}$. When the 
argument is small, the functions  $Y_1(x)$ and $K_1(x)$ are,
$${\pi\over 2}~Y_1(x)\approx -{1\over x} +\ln{x\over 2}\bigg({x\over 2}
+O(x^3)\bigg),~~~~K_1(x)\approx {1\over x} +\ln{x\over 2}\bigg({x\over 2}
+O(x^3)\bigg)~. $$
It is easy to see now that in the sum of the two integrals $dy$ over the
interval $(0,1)$, the leading singularities $dy/y$ exactly cancel each
other. Furthermore, it is safe to take the limits of $\xi\to 0$, $u\to 1$ in
the integrand, and even to set the lower limit of the integral $dq$ to be
zero. The resulting integral is convergent, $\xi$-independent, and yields a
term of  the order $\xi^2$. In the remaining integral, the variable $y$ runs
from $1$ to $\infty$, the integrand is not singular at finite $y$ and $q$,
and it is exponentially suppressed at large $y$ and $q$. The behavior of the
integral at $\xi\to 0$ is not singular and the basic upper estimate
$~const\times\xi^2$ remains unchanged for the entire integral
(\ref{eq:A2.04}).

\vspace{2cm}

\noindent {\bf ACKNOWLEDGMENTS}

The authors are grateful to Berndt Muller and Edward Shuryak for
helpful discussions at various stages in the development of this work,
and appreciate the help of Scott Payson who critically read the
manuscript.

\bigskip

\renewcommand{\theequation}{A\arabic{equation}}
\setcounter{equation}{0}

\section{Appendix. Wightman functions and propagators of wedge dynamics.}
\label{sec:SA1}

In this Appendix, we put all field correlators into a form which is needed
for the practical calculation of the self-energy. The density of states
$D_{[1]}$ and the causal part $D_{[0]}$ of the gluon propagator are used in
the form of decomposition over the transverse modes,
\begin{eqnarray}
  D_{[10]}^{lm}(\tau_2,\tau_1;\eta_2-\eta_1;\vec{k_t})= -i(2\pi)^2
  \int d\alpha \sum_{\lambda} v^{(\lambda)l}_{\alpha,\vec{k_t}}(\tau_2,\eta_2)
  \overstar{v}^{(\lambda)m}_{\alpha,\vec{k_t}}(\tau_1,\eta_1)~,\nonumber\\
 D_{[01]}^{lm}(\tau_2,\tau_1;\eta_2-\eta_1;\vec{k_t})= -i(2\pi)^2
  \int d\alpha \sum_{\lambda} v^{(\lambda)l}_{\alpha,-\vec{k_t}}(\tau_2,\eta_2)
  \overstar{v}^{(\lambda)m}_{\alpha,-\vec{k_t}}(\tau_1,\eta_1)~, 
\label{eq:A1.1}
\end{eqnarray}
where
\begin{eqnarray}  
v^{(TE)}_{{\vec k},\alpha}(x)={1\over 4\pi^{3/2} k_{t}} 
\left[ \begin{array}{c} 
                         k_y \\ 
                        -k_x \\ 
                         0 
                             \end{array} \right]
e^{-ik_{t}\tau\cosh (\alpha-\eta)};~~~
v^{(TM)}_{{\vec k},\alpha}(x)={1\over 4\pi^{3/2} k_{t}}  
\left[ \begin{array}{c} 
                 k_x f_1 \\ 
                 k_y f_1 \\ 
                  - f_2 
                 \end{array} \right] ~~,
\label{eq:A1.2}\end{eqnarray}  
are the transverse electric and transverse magnetic modes of the radiation 
field found previously in paper [II]. Here, we denoted,
\begin{eqnarray}  
f_1(\tau ,\eta)=  i \tanh (\alpha-\eta)
(e^{-ik_{t}\tau \cosh (\alpha-\eta)} - 1)~~, \nonumber\\ 
f_2(\tau ,\eta)= {e^{-ik_{t}\tau \cosh (\alpha-\eta)} -1 \over
\cosh^{2} (\alpha-\eta)} +
i k_{t}\tau {e^{-ik_{t}\tau\cosh (\alpha-\eta)} 
\over  \cosh (\alpha-\eta)}~~.
\label{eq:A1.3} 
\end{eqnarray}        
Starting from this form, we get the components of the commutator
$D_{[0]}(\tau_2,\tau_1;\eta_2-\eta_1;\vec{k_t})$, \footnote{In all formulae
below, the gluon rapidity $\alpha$ is counted from the reference point
$(\eta_1+\eta_2)/2$, the geometric center of the coordinates $\eta_1$ and
$\eta_2$ of the vertices in the self-energy loop. Thus, it corresponds to the
rapidity $\theta'$ in the integral representation of the quark correlators
in paper [II].}
\begin{eqnarray}
  D_{[0]}^{rs}=
\int {d\alpha\over 2\pi} \bigg\{\bigg[\delta_{rs}-{k_rk_s\over k_t^2}\bigg]
\sin k_tT_{12}
+{k_rk_s\over k_t^2}\tanh\big(\alpha+{\eta\over 2}\big) 
\tanh\big(\alpha-{\eta\over 2}\big)
\big[\sin k_tT_{12}\underline{-\sin k_tT_1 +\sin k_tT_2}~\big]\bigg\}~,
\label{eq:A1.4}
\end{eqnarray} 
\begin{eqnarray}
 D_{[0]}^{\eta\eta}=
\int {d\alpha\over 2\pi} 
{1\over k_t^2\cosh^2(\alpha+{\eta\over 2}) \cosh^2(\alpha-{\eta\over 2})}
\big[(1+k_t^2T_1T_2)\sin k_tT_{12}
-k_tT_{12}\cos k_tT_{12}\nonumber\\
\underline{+\sin k_tT_2-\sin k_tT_1-k_tT_2\cos k_tT_2 
+k_tT_1\cos k_tT_1}~\big]~,
\label{eq:A1.5}
\end{eqnarray} 
where $ T_1=\tau_1\cosh (\alpha-\eta/2)$, $ T_2=\tau_2\cosh (\alpha+\eta/2)$,
$T_{12}=T_1-T_2$. In the first of these equations, the coefficients of the
tensors  $(\delta_{rs}-k_rk_s/k_t^2)$ and $k_rk_s/k_t^2$ are the invariants 
${\cal D}^{(TE)}_{[0]}$ and ${\cal D}^{(2)}_{[0]}$ of Eq.~(\ref{eq:E3.8}), 
respectively. The latter is due to the $TM$-mode of the radiation field. Up to
the factor $k_t^{-2}$, Eq.~(\ref{eq:A1.5}) defines the invariant ${\cal
D}^{(\eta\eta)}_{[0]}$. The underlined terms are connected with the boundary
conditions imposed on the $TM$-mode at $\tau=0$. They cancel with the
underlined terms in the longitudinal part of the gauge field propagator 
given by Eqs.~(\ref{eq:A1.14}) and  (\ref{eq:A1.15}) thus
providing the causal behavior of the components $D_{[adv]}^{rs}$ and
$D_{[adv]}^{\eta\eta}$ of the retarded propagator $D_{[adv]}(\tau_2,\tau_1)$.
In the body of the paper, we call the residues of this cancelation as
the diagonal components  of $D^{[0]}$.
The ``off-diagonal'' components of the commutator $D_{[0]}^{ij}$ are
\begin{eqnarray}
D_{[0]}^{r\eta}= {-ik_r\over k_t^2}
\int {d\alpha\over 2\pi} 
{\tanh(\alpha+{\eta\over 2})\over\cosh^2(\alpha-{\eta\over 2})}
\big[\sin k_tT_{12}- k_tT_1\cos k_tT_{12} 
+\sin k_tT_2-\sin k_tT_1+k_tT_1\cos k_tT_1\big]~,
\label{eq:A1.6}
\end{eqnarray} 
\begin{eqnarray}
D_{[0]}^{\eta r}= {ik_r\over k_t^2}
\int {d\alpha\over 2\pi} 
{\tanh(\alpha-{\eta\over 2})\over\cosh^2(\alpha+{\eta\over 2})}
\big[\sin k_tT_{12}+ k_tT_2\cos k_tT_{12} 
+\sin k_tT_2-\sin k_tT_1-k_tT_2\cos k_tT_2\big]~.
\label{eq:A1.7}
\end{eqnarray}
The commutator is not a symmetric tensor. However, by examination, these
components are odd with respect to the rapidity  difference 
$\eta=\eta_1-\eta_2$, and hence they do not contribute to the effective quark
mass we are computing in this paper.

The tensor of the gluon density   $D_{[1]}^{ij}(\tau_2,\tau_1; \eta_2-\eta_1;
\vec{k_t})$ has the ``diagonal'' components,
\begin{eqnarray}
D_{[1]}^{rs}=
-i \int {d\alpha\over 2\pi} \bigg\{\bigg[\delta_{rs}-{k_rk_s\over k_t^2}\bigg]
\cos k_tT_{12} \hspace{8cm}\nonumber \\
+{k_rk_s\over k_t^2}\tanh(\alpha+{\eta\over 2}) \tanh(\alpha-{\eta\over 2})
(\cos k_tT_{12}-\cos k_tT_1 -\cos k_tT_2 +1)\bigg\}~,
\label{eq:A1.8}
\end{eqnarray} 
\begin{eqnarray}
 D_{[1]}^{\eta\eta}=
-i\int {d\alpha\over 2\pi} 
{1\over k_t^2\cosh^2(\alpha+{\eta\over 2}) \cosh^2(\alpha-{\eta\over 2})}
\big[(1+k_t^2T_1T_2)\cos k_tT_{12}
+k_tT_{12}\sin k_tT_{12}\nonumber\\
-\cos k_tT_2-\cos k_tT_1-k_tT_2\sin k_tT_2 -k_tT_1\sin k_tT_1 +1\big]~,
\label{eq:A1.9}
\end{eqnarray} 
from which one can infer the invariants ${\cal D}^{(TE)}_{[1]}$ and 
${\cal D}^{(2)}_{[1]}$ of Eq.~(\ref{eq:E3.8}) exactly in the same way as it
was done for the invariants of the commutator $D_{[0]}$. The off-diagonal
components,
\begin{eqnarray}
D_{[1]}^{r\eta}= {-k_r\over k_t^2}
\int {d\alpha\over 2\pi} 
{\tanh(\alpha+{\eta\over 2})\over\cosh^2(\alpha-{\eta\over 2})}
\big[\cos k_tT_{12}- k_tT_1\sin k_tT_{12} 
-\cos k_tT_2-\cos k_tT_1-k_tT_1\sin k_tT_1 +1\big]~,
\label{eq:A1.10}
\end{eqnarray} 
\begin{eqnarray}
D_{[1]}^{\eta r}= {k_r\over k_t^2}
\int {d\alpha\over 2\pi} 
{\tanh(\alpha-{\eta\over 2})\over\cosh^2(\alpha+{\eta\over 2})}
\big[\cos k_tT_{12}- k_tT_2\sin k_tT_{12} 
-\cos k_tT_2-\cos k_tT_1-k_tT_2\sin k_tT_2 +1\big]~,
\label{eq:A1.11}
\end{eqnarray}
are also non-symmetric and odd with respect to the rapidity  difference 
$\eta$. They also do not contribute to the effective quark mass. Equations
(\ref{eq:A1.8})-- (\ref{eq:A1.11}) give  the components of the vacuum density
of states of the gauge field in the wedge dynamics. In order to incorporate
the ``material'' part given by the distribution of real gluons, the integrand
of each of Eqs.~(\ref{eq:A1.8})-- (\ref{eq:A1.11}) must be multiplied by the
common factor $[1+2n_g(k_t,\alpha)]$.

The full tensor of the longitudinal part of the propagator that defines
the field $A(\tau_1)$ via the current $j(\tau_2)$ at all preceding times,
\begin{eqnarray}
A^{[long]}_l(\tau_1)=\int_{0}^{\tau_1}\tau_2~d\tau_2~d\eta_2
D^{[long]}_{lm}(\tau_2, \tau_1;\eta_1-\eta_2,{\vec k_t} ) j^m(\tau_2)~,
\label{eq:A1.12}
\end{eqnarray} 
was found in paper [III] in the following form,
\begin{eqnarray}
D^{[long]}_{lm}(\tau_2, \tau_1;\eta_1-\eta_2, {\vec k_t})
=\hspace{12cm}\\  \label{eq:A1.13}
=  \int {d\nu d^2{\vec k} \over (2\pi)^3 k_{\bot}^2}
\left[ \begin{array}{cc} 
 k_rk_s [Q_{-1,i\nu}(k_{\bot}\tau_2)- Q_{-1,i\nu}(k_{\bot}\tau_1)] &
 k_r \nu [ Q_{1,i\nu}(k_{\bot}\tau_2) -Q_{-1,i\nu}(k_{\bot}\tau_1)]\\
 \nu k_s  [ Q_{-1,i\nu}(k_{\bot}\tau_2) -Q_{1,i\nu}(k_{\bot}\tau_1)] &
 \nu^2  [ Q_{1,i\nu}(k_{\bot}\tau_2) -Q_{1,i\nu}(k_{\bot}\tau_1)]
                                            \end{array} \right]_{lm} 
 e^{-i\nu(\eta_1-\eta_2) }~.\nonumber     
\end{eqnarray}  
The diagonal components of this longitudinal part of the gluon propagator 
are just the differences between the vector potentials of the ``static''
gluon fields at the final time $\tau_1$ and the initial time $\tau_2$,
\begin{eqnarray}
 D^{[long]}_{rs}= {k_rk_s\over k_t^2}
 \bigg\{ {\coth |\eta|\over 2} (e^{-\tau_1 k_t\sinh |\eta|}-
 e^{-\tau_2 k_t\sinh |\eta|}) \hspace{6cm}\nonumber \\
 -~ \underline{\int {d\alpha\over 2\pi}
 \tanh(\alpha+{\eta\over 2}) \tanh(\alpha-{\eta\over 2})
\big[~\sin k_tT_1-\sin k_tT_2~\big]}~\bigg\}~,
\label{eq:A1.14}
\end{eqnarray} 
\begin{eqnarray}
 D^{[long]}_{\eta\eta}= -  {\tau_1^2-\tau_2^2\over 2}\delta(\eta)
 -\bigg\{\bigg[\bigg({\cosh\eta\over k_t^2\sinh^3|\eta|}+
 {\tau_1\cosh\eta\over k_t\sinh^2\eta}+
 {\tau_1^2\cosh\eta\over 2\sinh |\eta|}\bigg)e^{-\tau_1 k_t\sinh |\eta|}\bigg]
 -\bigg[\tau_1\to\tau_2\bigg]\bigg\} \nonumber\\
 -\underline{\int {d\alpha\over 2\pi}{1\over k_t^2\cosh^2(\alpha+{\eta\over 2})
 \cosh^2(\alpha-{\eta\over 2})}
\big[~  \sin k_tT_1-\sin k_tT_2-k_tT_1\cos k_tT_1 
 +k_tT_2\cos k_tT_2~\big]} ~.
\label{eq:A1.15}
\end{eqnarray} 
By the derivation, these components include $\theta(\tau_1-\tau_2)$ of the
following origin. The source current  which acts at the moment $\tau_2$
produces the  simultaneous longitudinal electric field $E(\tau_2)$. The gauge
field potential is rebuilt from the electric field at the time $\tau_1>\tau_2$
by integrating the electric field $E(\tau)$ over all times from $0$ to
$\tau_1$. The underlined terms in Eqs.~(\ref{eq:A1.14}) and (\ref{eq:A1.15})
cancel out in the full assembly of the retarded propagator
$D_{[adv]}(\tau_2,\tau_1)$ with the underlined terms in the radiation part,
Eqs.~(\ref{eq:A1.4}) and (\ref{eq:A1.5}). In the body of the paper, we
call  the residue of this cancelation as the diagonal components of 
$D^{[long]}$, which can be conveniently written as
\begin{eqnarray}
 D^{[long]}_{rs}= - {k_rk_s\over k_t^2}
 ~{k_t\cosh \eta\over 2} 
 \int_{\tau_2}^{\tau_1}  e^{-t k_t\sinh |\eta|} dt~,
\label{eq:A1.14a}
\end{eqnarray} 
\begin{eqnarray}
 D^{[long]}_{\eta\eta}= -  {\tau_1^2-\tau_2^2\over 2}\delta(\eta)
 + {k_t\cosh \eta\over 2} 
 \int_{\tau_2}^{\tau_1}  e^{-t k_t\sinh |\eta|} t^2~ dt~.
\label{eq:A1.15a}
\end{eqnarray} 
Once again, the ``off-diagonal'' components of the longitudinal part of
propagator,
\begin{eqnarray}
D^{[long]}_{r\eta}= {i k_r\over k_t^2}~
 \big\{ -{{\rm sign}\eta\over 2\sinh^2\eta} (e^{-\tau_1 k_t\sinh |\eta|}-
 e^{-\tau_2 k_t\sinh |\eta|}) 
 +{k_t\tau_2\over\sinh\eta}e^{-\tau_2 k_t\sinh |\eta|}-
 {k_t\tau_1\cosh^2\eta\over\sinh\eta}e^{-\tau_1 k_t\sinh |\eta|}\nonumber\\
 -\int {d\alpha\over 2\pi}
 {\tanh(\alpha-{\eta\over 2})\over  \cosh^2(\alpha+{\eta\over 2})}
 [\sin k_tT_1-\sin k_tT_2+ k_t T_2\cos k_t T_2]\big\}
\label{eq:A1.16}
\end{eqnarray} 
\begin{eqnarray}
D^{[long]}_{\eta r}= {i k_r\over k_t^2}
 \big\{ -{{\rm sign}\eta\over 2\sinh^2\eta} (e^{-\tau_1 k_t\sinh |\eta|}-
 e^{-\tau_2 k_t\sinh |\eta|}) 
 -{k_t\tau_1\over\sinh\eta}e^{-\tau_1 k_t\sinh |\eta|}
 +{k_t\tau_2\cosh^2\eta\over\sinh\eta}e^{-\tau_2 k_t\sinh |\eta|}\nonumber\\
 -\int {d\alpha\over 2\pi}
 {\tanh(\alpha+{\eta\over 2})\over  \cosh^2(\alpha-{\eta\over 2})}
 [-\sin k_tT_1+\sin k_tT_2+ k_t T_1\cos k_t T_1]\big\} 
\label{eq:A1.17}
\end{eqnarray}
are odd with respect to $\eta$ and do not contribute the effective quark mass.

The gauge-field correlators have several distinctive features. First,  the
lengthy expression for each component is such that the gauge field correlators
obey the boundary condition  $A_{\eta}(\tau =0, \vec{r_t})=0$  which provides
continuity of the field at $\tau=0$, and allows for a complete fixing of the
gauge. Second, in the $rs-$ and $\eta\eta$-components of the propagator 
\begin{eqnarray}
D^{lm}_{[adv]}(\tau_2,\tau_1,\eta;k_t)=
-\theta(\tau_1-\tau_2)D^{lm}_{[0]}(\tau_2,\tau_1,\eta;k_t)
+D^{lm}_{L}(\tau_2,\tau_1,\eta;k_t)~,
\label{eq:A1.18}
\end{eqnarray}
the boundary terms cancel between the transverse and longitudinal parts.
This fact provides causal behavior of the $\Sigma(\tau_1,\tau_2)$ that
defines the dispersion law.

The fermion invariants $g_{[\alpha]}$ were derived in paper [II].
For the sake of completeness, we reproduce the final answers here,
\begin{eqnarray}
 g^{L(\pm)}_{[0]}=
 i~{\tau_1 e^{\mp\eta/2}-\tau_2e^{\pm\eta/2}\over 4\sqrt{|\tau_{12}^2|}}
 \theta(\tau_{12}^2)~J_1(q_t\sqrt{|\tau_{12}^2|})~,~~~~
 g^{T(\pm)}_{[0]}= -~{ e^{\mp\eta/2}\over 4}
 \theta(\tau_{12}^2)~J_0(q_t\sqrt{|\tau_{12}^2|})~,
 \label{eq:A1.19}\end{eqnarray}
  \begin{eqnarray}
  g^{L(\pm)}_{[1]}=  - \int{d\theta' \over 4\pi} 
\big[1- 2 n_f\big({\eta_1+\eta_2\over 2}+ \theta',p_t\big)\big]  
~e^{\mp\theta'}
~\sin\big( p_t[\tau_1\cosh(\theta-\eta/2)- 
   \tau_2\cosh(\theta+\eta/2)]\big)\nonumber\\
={\tau_1 e^{\mp\eta/2}-\tau_2e^{\pm\eta/2}\over 4\sqrt{|\tau_{12}^2|}}
  \bigg[\theta(\tau_{12}^2)~Y_1\bigg(q_t\sqrt{|\tau_{12}^2|}\bigg)
  +{2\over\pi}\theta(-\tau_{12}^2)
  K_1\bigg(q_t\sqrt{|\tau_{12}^2|}\bigg)\bigg] 
 ~\big[ 1-2n_f(q_t) \big] ~,\nonumber\\
g^{T(\pm)}_{[1]}= -i e^{\mp\eta/2} \int{d\theta' \over 4\pi}
\big[1- 2 n_f\big({\eta_1+\eta_2\over 2}+ \theta',p_t\big)\big]
~\cos\big( p_t[\tau_1\cosh(\theta-\eta/2)- 
   \tau_2\cosh(\theta+\eta/2)]\big)\nonumber\\
= i{ e^{\mp\eta/2}\over 4}
  \bigg[\theta(\tau_{12}^2)~Y_0\bigg(q_t\sqrt{|\tau_{12}^2|}\bigg)
  -{2\over\pi}\theta(-\tau_{12}^2)
  K_0\bigg(q_t\sqrt{|\tau_{12}^2|}\bigg)\bigg]~\big[ 1-2n_f(q_t) \big]~.
\label{eq:A1.20}
\end{eqnarray}

\end{document}